\documentclass[aps,prd,twocolumn,showpacs,preprintnumbers,superscriptaddress,amsfont,graphicx,nofootinbib]{revtex4-1}
\usepackage[utf8]{inputenc}
\usepackage{amsmath}
\usepackage{amsthm}
\usepackage{amssymb}
\usepackage{enumitem}   
\usepackage[english]{babel}
\usepackage{url}
\usepackage{mathtools}
\usepackage{bbold}
\usepackage{slashed}
\usepackage{multirow}
\usepackage{xcolor}
\usepackage{slashed}
\usepackage{multirow}
\usepackage[toc,page]{appendix}
\usepackage{hyperref}

\usepackage{amsmath, amsfonts, amssymb}
\usepackage{graphicx}
\usepackage{color}
\usepackage{enumerate}
\usepackage{hyperref}
\usepackage{latexsym}
\usepackage{dsfont}
\usepackage{hepnicenames}
\usepackage{enumerate}
\usepackage{soul}
\usepackage[normalem]{ulem}
\usepackage{bbold}
\usepackage{wasysym}
\usepackage{makecell}
\usepackage{slashed} 
\usepackage{comment,verbatim}
\usepackage{amsmath}

\usepackage{graphicx,amsmath,amssymb}
\usepackage{color,epsfig}
\usepackage{comment}
\usepackage{xcolor}
\usepackage{graphicx,color}
\def\be{\begin{equation}}
\def\ee{\end{equation}}

\begin{document}

\title{
Domain walls 
seeding the electroweak
phase transition}

\author{Simone Blasi} 
\email{simone.blasi@vub.be}
\author{Alberto Mariotti}
\email{alberto.mariotti@vub.be}
\affiliation{Theoretische Natuurkunde and IIHE/ELEM, Vrije Universiteit Brussel, \& The  International Solvay Institutes, Pleinlaan 2, B-1050 Brussels, Belgium}

\begin{abstract}
Topological defects can act as local impurities that seed
cosmological phase transitions. In this paper we study the case
of domain walls, and how they can
affect the electroweak phase transition
in the Standard Model extended with a $Z_2$--odd scalar
singlet.
When the transition is two--step, the early breaking of the $Z_2$
symmetry implies the formation of domain walls which 
can then act as nucleation sites for the second step.
We develop a method based on dimensional reduction to calculate
the rate of the catalyzed phase transition within the 3d theory on the domain wall surface. 
By comparison with the standard
homogeneous rate, we conclude that 
the seeded phase transition 
is generically faster and it ultimately determines the way the phase transition
is completed. We comment
on the phenomenological implications
for gravitational waves and baryogenesis.
\end{abstract}

\maketitle

\tableofcontents

\section{Introduction}
\label{sec:int}

The study of
cosmological phase transitions 
is of great importance in high energy physics 
as it can shed light on major open questions 
in the Standard Model (SM), such as the origin
of the matter--anti matter
asymmetry in the Universe,
and can help us uncovering the dynamics
of the electroweak symmetry breaking
thanks to the exciting prospects
for detecting the corresponding background of 
gravitational waves in the near 
future\,\cite{Barausse:2020rsu,Caldwell:2022qsj}.

The physics of first order phase transitions  
crucially depends on the mechanism controlling
the bubble nucleation.
This is usually assumed to proceed 
via thermal or quantum fluctuations in a 
homogeneous spacetime. There is, however,
the possibility that inpurities
that are present in the Universe at the time of the phase
transition can actually seed the bubble nucleation,
thus providing a competing mechanism to complete
the phase transition\,\cite{Steinhardt:1981ec,
Steinhardt:1981mm,Jensen:1982jv,Hosotani:1982ii,
PhysRevD.30.272}.
Following \cite{Steinhardt:1981ec}, we will refer to the first case as 
homogeneous nucleation,
and to the second case as inhomogeneous 
or seeded nucleation, since in the latter
the tunneling probability is not uniformly distributed on the spacetime,
but it is enhanced at the location of the seeds.

The nature of the impurities can be vastly
different, ranging from black holes
\,\cite{Hiscock:1987hn,Green:2006nv,Gregory:2013hja,Burda:2015yfa,Mukaida:2017bgd,Mukaida:2017bgd,El-Menoufi:2020ron}
and local over--densities\cite{Oshita:2018ptr,Balkin:2021zfd},
to topological defects \cite{Steinhardt:1981ec,
Steinhardt:1981mm,Jensen:1982jv,
PhysRevD.30.272,Yajnik:1986tg,Yajnik:1986wq,Preskill:1992ck,Dasgupta:1997kn,
Kumar:2008jb,Lee:2013zca,Kumar:2009pr,
Kumar:2010mv,Lee:2013ega,Koga:2019mee,Dunsky:2021tih,Agrawal:2022hnf},
with also possible implications for the lifetime of the metastable 
SM vacuum \cite{Isidori:2001bm,Degrassi:2012ry,Buttazzo:2013uya}.
Concerning phase transitions seeded by topological defects, 
previous studies have mostly
focussed on cosmic strings\,\cite{Yajnik:1986tg,Yajnik:1986wq,Dasgupta:1997kn,
Kumar:2008jb,Lee:2013zca,Koga:2019mee} 
and monopoles\,\cite{Steinhardt:1981ec,Kumar:2009pr,Kumar:2010mv,Agrawal:2022hnf}.

In this paper we investigate the case of domain walls,
topologically stable two--dimensional objects
forming at the spontaneous breakdown of a 
discrete symmetry\,\cite{Zeldovich:1974uw,Kibble:1976sj}.
In particular, we will focus on two--step phase transitions
where the defects created in the first step can act as seeds for the second one.
As a primary case of study,
we consider the possibility
that domain walls have played a role in the electroweak
phase transition. 

As a matter of fact, probably the simplest
physics beyond the SM that can turn the electroweak
crossover into a first order phase transition is the
addition of a real scalar field, $S$, singlet under the 
SM gauge group\,\cite{McDonald:1993ex,Burgess:2000yq,Espinosa:2007qk,Profumo:2007wc,Barger:2007im,Espinosa:2008kw,Espinosa:2011ax,Cline:2012hg,Profumo:2014opa,Feng:2014vea,
Curtin:2014jma,Craig:2014lda,Huang:2016cjm,Vaskonen:2016yiu,Curtin:2016urg,Kurup:2017dzf,Buttazzo:2018qqp,Caprini:2019egz,Alanne:2019bsm}.
Since our focus is the role of domain walls,
we will be interested in the case in which $S$
is odd under a $Z_2$ symmetry. 
This scenario actually implies a very minimal set of new physics effects, 
given that any mixing between the Higgs and the singlet
is suppressed, and it has been studied
extensively at present and future 
colliders\,\cite{Curtin:2014jma,Craig:2014lda,Huang:2016cjm,Buttazzo:2018qqp,Costantini:2020stv,AlAli:2021let}.

A typical way in which the phase transition
can proceed in this scenario is indeed by a two--step
process in which first the singlet $S$ develops
a non--zero vacuum expectation value (vev),
and a second step (which is usually first order) 
leading to the breakdown of the electroweak symmetry.
Given that the singlet is odd under a $Z_2$, the first step 
will lead to the formation of domains whose vacua are related
by this symmetry, with walls forming at their boundaries according
to the Kibble mechanism\,\cite{Kibble:1976sj}. 

These domain walls
are however not cosmologically stable if the actual
vacuum of the theory entails a vanishing singlet vev, so that the 
$Z_2$ symmetry is eventually restored at the end
of the electroweak phase transition, as also noted in Ref.\,\cite{Espinosa:2011eu}. 
This means that domain walls can act as seeds
for the phase transition while being
collapsed upon its completion.

The purpose of this paper is to study the details
of this process.
As we shall see, domain walls can catalyze
the electroweak phase transition in two ways, namely
either by developing a classical instability and dissociating,
or by acting as nucleation sites
favoring the quantum/thermal tunneling.
In order to calculate the rate of this tunneling
we will develop a formalism based on a 
Kaluza-Klein reduction technique,
which could in principle be extended to study other defects as well.
We will also present a complementary approach based on the
thin--wall approximation to gain further
intuition on this inhomogeneous nucleation.

In the parameter space that we have
covered for the Higgs-singlet model, we find that the 
inhomogeneous transition is always faster than the homogeneous
one. 
Moreover, 
regions of parameter space in which bubbles
would fail to nucleate according to the standard 
tunnelling rate can now become viable thanks to the domain
walls--- see Fig.\,\ref{fig:scan180} for an overview
of our results.
This can largely
affect the phenomenological properties of the electroweak phase
transition,
such as for instance its duration 
and the geometry of the nucleated bubbles
with implications for the gravitational
wave signal,
as discussed in Sec.\,\ref{sec:discussion}.
together with interesting future directions.

\section{Seeded vs homogeneous vacuum decay}
\label{sec:svsuns}
The phase transition seeded by domain walls 
can proceed via two mechanisms:
\begin{itemize}
 \item domain walls can become classically
unstable while the Universe cools down, developing
a region of the true vacuum in their interior, 
and then dissociate;
\item even if an instability never develops, 
domain walls can nevertheless facilitate the tunneling
towards the true vacuum by acting as local impurities.
\end{itemize}
Both these possibilities, dubbed \emph{Rolling} and \emph{Tunneling}, respectively,
can be described in the dimensionally--reduced 
theory presented in Sec.\,\ref{sec:formalism}.

In order to first understand in generality the dynamics of the 
seeded tunneling, 
we here revisit the computation of the nucleation condition
for the homogeneous phase transition and compare it with the one derived when 
domain walls act as nucleation sites.

The rate of false vacuum decay per unit volume for a homogeneous tunneling
can be written 
as\,\cite{PhysRevD.15.2929,Linde:1980tt,Linde:1981zj}
\begin{equation}
 \gamma_V \equiv \frac{\Gamma}{V} = A \,\text{exp}(-S_3/T),
\end{equation}
where $S_3$ is the time--independent euclidean action
evaluated on the $O(3)$ symmetric bounce solution, and the prefactor 
can be estimated as $A \sim T^4$. 
In order to determine the nucleation
temperature, $T_n$, at which the phase transition
actually takes place in an expanding Universe, one needs to impose
(see e.g.\,\cite{Ellis:2018mja})
\begin{equation}
\mathcal{N}(T_n) = \int_{T_n}^{T_c} \frac{\gamma_V}{H^4} \frac{\text d T}{T}
=1
\end{equation}
where $\mathcal{N}$ is the number of bubbles nucleated inside
one Hubble volume, $H^{-3}$, in the range between the critical 
temperature, $T_c$, and $T_n$. The integral for $\mathcal{N}$ is typically
dominated by temperatures around $T_n$. Moreover, with the notable
exception of strongly supercooled phase transitions,
we can assume radiation domination throughout the whole nucleation
process. This leads to the standard condition
\begin{equation}
\label{eq:nucle_homo}
 \frac{S_3(T_n)}{T_n} \simeq 4 \,\text{log}(M_\text{Pl}/T_n) - 11.4
 \approx 145,
\end{equation}
where in the last step we have set $T_n$ around the electroweak scale,
and $M_\text{Pl}$ is the Planck mass.

Let us now move to discuss the case of a tunneling process
than can only occur in the vicinity of an impurity,
which we shall take to be a domain wall. 
It is then natural to define a nucleation rate per unit surface
given by\,\cite{Preskill:1992ck,Dasgupta:1997kn}
\begin{equation}
 \gamma_S \equiv \frac{\Gamma}{S} = A^\prime \,\text{exp}(-S_\text{inh})
\end{equation}
where $A^\prime \sim \sigma_\text{DW}$ is the tension of the 
domain wall
and $S_\text{inh}$ is the action describing the inhomogeneous
tunneling.
In order to evaluate the nucleation condition 
we parameterize the total surface occupied by domain walls
inside one Hubble volume as $S_H = \xi H^{-2}$,
where $\xi$ is $\mathcal{O}(1)$ in the scaling regime \cite{Vilenkin:2000jqa}.
The new condition then reads 
\begin{equation}
\mathcal{N}(T_n) =  \int_{T_n}^{T_c} \xi \frac{\gamma_S}{H^3} \frac{\text d T}{T}
=1
\end{equation}
and therefore
\begin{equation}
\label{eq:nucDW}
 S_\text{inh} = 3 \, \text{log}(M_\text{Pl}/T_n) + 
 \text{log}(\sigma_{\text{DW}}/T_n^3)+
 \text{log}\, \xi
 - 8.5 \approx 105
\end{equation}
for a transition around the electroweak scale, where we have assumed
in addition that also the tension of the domain wall
is set by the electroweak scale, as this will be the focus
of our study.

The bounce action $S_\text{inh}$ is
associated with the formation of a bubble that modifies the
original domain wall profile,
and that also extends in the $z$ direction orthogonal
to it. Its geometry
will then deviate from the $O(3)$ spherical symmetry
at high temperatures, but it will preserve
the $O(2)$ rotational invariance in the plane
parallel to the domain wall described
by $r$.
As we shall see in Sec.\,\ref{sec:bubble},
the shape of the bubble may 
then resemble the one of a spheroid 
nucleated inside the domain wall.

The euclidean action setting the tunneling
rate can be computed once the field profiles corresponding to the bounce solution
are known.
In the case of a tunneling seeded by domain walls,
we may then search
for a high--temperature solution for the fields $\Phi_i$ 
involved in the tunneling process,
$\Phi_i = (S,h)$
for the Higgs--singlet model,
that is independent
of the euclidean time, $\Phi_i(r,z)$.
The boundary conditions will be such
that the unperturbed domain wall is recovered for
$r\rightarrow \infty$ and $z\rightarrow \pm \infty$,
together with $\partial_r \Phi_i(r=0,z)=0$. The 
bounce action at the temperature $T$ will then be given
in this case by $S_\text{inh}= S_2/T$, where the subscript
refers to the $O(2)$ symmetry of the solution.

This solution may be obtained
by solving the coupled system of partial differential
equations with the aforementioned boundary conditions.
In Sec.\,\ref{sec:KK} we will instead introduce a new approach
based on a Kaluza--Klein (KK) decomposition along the direction
orthogonal to the domain wall. 
As we shall see, this provides an alternative
way of computing the bounce action (and hence the transition rate),
and at the same time gives a more physical picture 
of the inhomogeneous tunneling.

\section{The SM plus scalar singlet}
\label{sec:setup}

Before introducing the formalism needed to describe
the seeded phase transition, we shall here introduce 
our case of study, namely the SM extended with a real scalar 
singlet that is odd under a $Z_2$ symmetry.
The potential at the renormalizable level in the
unitary gauge reads
\begin{equation}\label{eq:V4d}
 V(h,S) = -\frac{\mu^2}{2} h^2 + 
 \frac{\lambda}{4} h^4 -\frac{m^2}{2} S^2
 +\frac{\eta}{4} S^4 + \frac{\kappa}{2} h^2 S^2,
\end{equation}
where the Higgs component $h$ is such that $\mathcal{H} = (0,h)/\sqrt{2}$,
and we shall take
$\kappa >0$ and $\mu^2, m^2 >0$.
The parameter $m^2$ can be traded for the physical singlet mass
in the $(h=v,S=0)$ vacuum at zero temperature, $m_S^2$, as
\begin{equation}\label{eq:ms}
 m^2 = \kappa v^2 - m_S^2.
\end{equation}

Since our focus is the seeded vacuum decay,
we will only consider the case in which 
the electroweak phase transition occurs in two steps,
as we shall briefly discuss.
Assuming that the vacuum at high temperatures
corresponds to vanishing $h$ and $S$ so that 
all symmetries are ``restored'', the first step entails
the spontaneous breaking of the discrete $Z_2$ symmetry
at the temperature
$T_\text{d}$, with the singlet scalar developing
a non--vanishing vacuum expectation value (vev)
$v_s = m/\sqrt{\eta}$, while the 
Higgs minimum is still at zero.
At this point, given that $\langle S \rangle = \pm v_s$
are energetically equivalent configurations,
regions with $+$ and $-$ will form inside a Hubble patch
separated by domain walls at the boundaries, with profile given by
\begin{equation}\label{eq:DW}
 S_\text{DW}(z) = v_s \,\text{tanh}(m z/\sqrt{2}).
\end{equation}
At temperatures below the critical temperature, $T_c$,
the $(0,\pm v_s)$ vacuum becomes metastable and decays
to the (true) electroweak vacuum $(v,0)$.
The two--step phase transition can then be summarized as
\begin{equation}\label{eq:twostep}
 (0,0) \rightarrow (0, v_s) \rightarrow (v,0).
\end{equation}
The second step of the chain is usually
assumed to occur in homogeneous spacetime, equivalently
either inside the $+$ or $-$ domains,
and it is typically first--order.
However, if walls are present at the time of electroweak
symmetry breaking they can induce a seeded vacuum decay
that can only take place close to the boundary
between the domains. 
In fact, at the center of the domain wall, $z=0$ in 
\eqref{eq:DW}, the singlet vev is vanishing, and so
would be the barrier 
to reach the
true vacuum
according to the potential in \eqref{eq:V4d}.
Even though an actual barrier still exists due to
the non--zero gradient of the domain wall profile, this argument
gives an intuitive understanding of how 
the phase transition can be facilitated by the
domain walls.
This process then
provides a competing mechanism for the second
step in \eqref{eq:twostep},
since homogeneous 
tunneling can of course still occur far away from the domain walls.

In order to have a seeded vacuum
decay, domain walls formed at $T_\text{d}$ need
not to collapse before the electroweak phase transition
has completed. This aspect is connected to the quality
of the corresponding discrete symmetry. 
In fact, the effect of small $Z_2$--breaking terms in the 
scalar potential \eqref{eq:V4d} is to induce a bias $\epsilon$ between 
the two minima at $\pm v_s$, 
which acts as a vacuum pressure that tends to 
expand the domain corresponding to the deepest minimum
over the other until the domain structure
is destroyed.
This collapse can of course be
prevented by the domain wall tension.
However, if the characteristic scale
of the network (and hence
the curvature radius) is of order $H^{-1}$
as eventually expected in the scaling regime, the collapse
will be almost 
instantaneous after formation,
unless the energy difference $\epsilon$ is extremely
small, see e.g. \,\cite{McDonald:1995hp,Espinosa:2011eu}.
In practice, this means that the $Z_2$ needs to be exact for
all the operators up to dimension four.

On the other hand, no ``special'' quality for the $Z_2$
symmetry is required if its breaking is due to
gravity. In fact, even if this breaking is
generated already at dimension five with an operator
$S^5/\Lambda$, domain walls can still play a role in
the electroweak phase transition
if the cutoff $\Lambda$ is at the Planck scale
and $v_s \sim 100$ GeV\,\cite{McDonald:1995hp}.
Indeed the domain wall annihilation temperature, $T_\text{ann}$, can be estimated
by requiring that the tension force $\sigma_\text{DW}/R$
(where $\sigma_\text{DW} \sim \eta^{1/2} v_s^3$ is the domain wall tension
and $R \sim 1/\xi H$ is the typical curvature radius)
equals the vacuum pressure due to the bias,
$\epsilon \sim c \, v_s^5/M_\text{Pl}$ with $c$ a $\mathcal{O}(1)$
coefficient.
We then obtain 
\begin{equation}\label{eq:Tann}
 \frac{T_\text{ann}}{v_s} \sim 0.5 
 \frac{c^{1/2}}{\xi^{1/2} \eta^{1/4}},
\end{equation}
where the Planck scale drops out from the calculation
as this sets both the typical curvature radius through
Hubble and the size of the bias.
We then conclude that a $Z_2$ explicit breaking
due to gravity is generically not enough
to make the domain walls collapse before the 
electroweak phase transition,
as $T_\text{ann}$ in \eqref{eq:Tann}
can easily be below the nucleation temperature.

We nevertheless note that 
the model can have a viable phenomenology also with
an exact $Z_2$ symmetry:
even though scalar singlet dark matter with large portal
coupling to the Higgs boson is almost excluded
for the mass range of interest, $S$ needs not to be
the lightest state charged under this symmetry.
In addition, domain walls are anyways collapsed
at the end of the electroweak phase transition
ensuring no contradiction with standard cosmology.

\section{Formalism for the seeded phase transition}
\label{sec:formalism}

The seeded phase transition can be described 
as a process in which the unperturbed domain wall
is modified by the onset of a Higgs vev at its core
eventually filling up the whole space. 
As mentioned in Sec.\,\ref{sec:svsuns}, 
this can occur either via a classical
instability of the unperturbed domain wall profile (rolling), or,
if this configuration is metastable, via thermal or
quantum tunneling.

Both these scenarios can be captured by analyzing 
the lower--dimensional effective theory obtained 
by performing a KK decomposition
along the direction orthogonal to the domain wall.
This strategy is discussed in Sec.\,\ref{sec:KK}.
In addition, we can gain further physical intuition
by resorting to a thin wall approximation in which
the role of the pre--exhisting domain wall becomes manifest
in facilitating the tunneling to the true vacuum. This will
be the topic of Sec.\,\ref{sec:TW}.

Interestingly,
these two formalisms can actually describe the tunneling
process in complementary temperature ranges,
namely far from $T_c$ (KK decomposition), and close
to $T_c$ (thin wall),
as we shall discuss in detail in Sec.\,\ref{sec:results}.

\subsection{Kaluza--Klein decomposition}
\label{sec:KK}

The starting point is to make an
ansatz for the singlet and the Higgs field in the background
of the unperturbed domain wall as
\begin{equation}\label{eq:ansatz}
S = S_\text{DW}(z) + \sum_k s_k(x) \sigma_k(z),
\,\,
h = \sum_k h_k(x) \phi_k(z)
\end{equation}
where $S_\text{DW}(z)$ is the profile given
in Eq.\,\eqref{eq:DW}, which is a solution to the equations
of motion together with $h=0$, and we have denoted
by $x$ the remaining three spacetime coordinates.
The sum runs over a complete set of eigenfunctions,
$\sigma_k(z)$ and $\phi_k(z)$.
The ansatz \eqref{eq:ansatz} can be plugged into
the 4d action of the Higgs--singlet model
in order to obtain a theory for the 3d modes $s_k(x)$ and $h_k(x)$,
which will allow us to study the occurrence of classical
instabilities, and furthermore to calculate the seeded
tunneling rate as a standard homogeneous process in three
spacetime dimensions.

The 4d potential \eqref{eq:V4d} is of course
temperature dependent. This means that in principle a new decomposition
needs to be performed at each temperature of
interest. As we shall see, however, in some limits
such as the leading high temperature expansion this is not
necessary. We will then first present our procedure at 
$T=0$ and then discuss a simple way to extend it at finite temperature.

\begin{figure}
\hspace{-10mm}
 \includegraphics[scale=0.53]{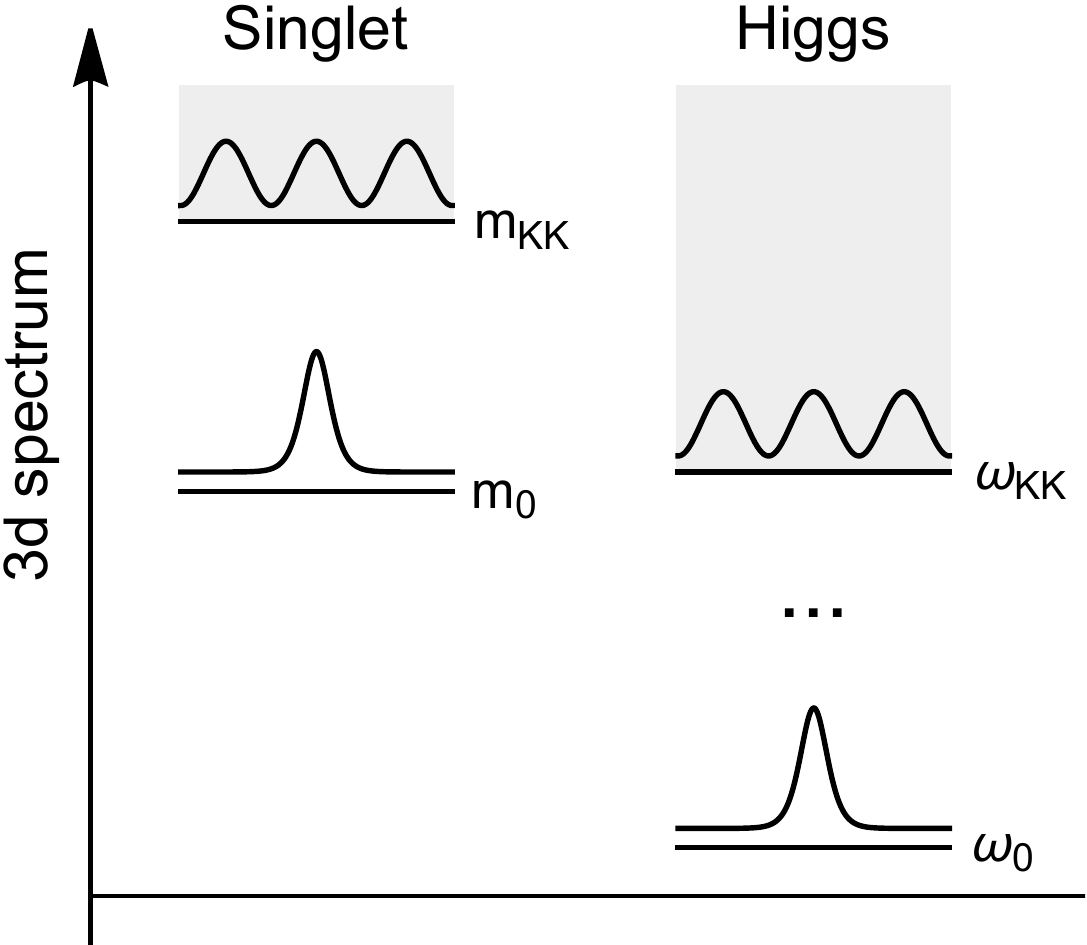}
 \caption{
A cartoon of the spectrum of the 3d theory according to 
the eigenvalue equations \eqref{eq:sigmak} and \eqref{eq:linhiggs}.
The singlet modes consist of a single bound state (neglecting
the massless mode from the breaking of translational invariance)
with mass $m_0^2 >0$ indicated by the solid line, together
with a sketch of its localized profile. On top of this 
there is a gapped continuum of KK states shown by 
the upper gray region, with plane--wave
asymptotic behaviour.
The Higgs spectrum is qualitatively the same besides the fact that
more discrete modes can in principle be present,
and that the mass of the lightest state $\omega_0^2$ 
needs not to be positive.
\label{fig:KKcartoon}
 } 
\end{figure}

In order to diagonalize the quadratic part of the 3d action,
the profiles $\sigma_k$ and $\phi_k$ are taken to be solutions
to the eigenvalue equations
\begin{align}
\label{eq:sigmak}
& -\sigma^{\prime\prime}_k +
 (3\eta S_\text{DW}^2(z) -m^2)\sigma_k
 = m_k^2 \sigma_k, \\
 \label{eq:linhiggs}
&  -\phi^{\prime\prime}_k + (\kappa S_\text{DW}^2(z) -\mu^2)\phi_k
 = \omega_k^2 \phi_k,
\end{align}
which correspond to a P\"osch--Teller potential
and can be solved exactly in terms of Legendre
polynomials. A cartoon of the spectrum for the Higgs
and singlet states is shown in Fig.\,\ref{fig:KKcartoon}.

The singlet spectrum 
consists of a discrete mode, $\sigma_0$, corresponding
to a bound state,
and a tower of scattering states, $\sigma_k$,
with plane--wave asymptotic behaviour. The corresponding
masses are given by (see e.g.\,\cite{Rajaraman:1982is})
\begin{equation}\label{eq:sspec}
 m^2_0 = \frac{3}{2} m^2,
 \quad m^2_k= \left( \frac{q_k^2}{2} + 2\right) m^2
\end{equation}
and we have neglected the massless mode associated to the 
breaking of translational invariance as it will play no role 
in what follows. Here $q_k>0$ denote the dimensionless
wavevectors of the continuum states which for concreteness
we can consider normalized in a box of size $z\in(-L,L)$.
The corresponding eigenfunctions read
\begin{align}
& \sigma_0(z) \propto \frac{\text{tanh}(b z)}{\text{cosh}(b z)}, \\
& \sigma_k(z) \propto e^{i q_k b z}(3 \,\text{tanh}^2(b z) -1 - q^2_k
 - 3 i q_k \,\text{tanh}(b z) ),
 \nonumber
\end{align}
where we have introduced $b=m/\sqrt{2}$.

The spectrum of the Higgs states
is qualitatively the same. 
It consists of $n$ discrete modes (bound states) with masses
$\omega^2_n$ given by
\begin{equation}\label{eq:omegap}
 \omega^2_n
 = \frac{m^2}{2}
\left( p + 2np -n^2 \right)
-\mu^2,
 \end{equation}
where $n=p - \nu$ with $p\,(p+1)/2=\kappa/\eta$, namely
\begin{equation}
 p = \frac{1}{2} \left( \sqrt{1+\frac{8 \kappa}{\eta}}-1\right).
\end{equation}
The allowed values for $\nu$ are 
\begin{equation}
0 < \nu = p, \,p-1, \,p-2, \dots
\end{equation}
and the localized profiles are given by
\begin{equation}
 \phi_n(z) \propto
 (1-x^2)^{-\nu/2}
 \frac{\text{d}^n}{\text{d}x^n}(1-x^2)^p\big|_{x = \,\text{tanh}\,\frac{m z}{\sqrt{2}}},
\end{equation}
with $\mathcal{O}(m^{-1})$ spread around $z=0$.
When $n$ is zero or even, the profile will be symmetric
for $z\rightarrow-z$, and antisymmetric otherwise. 
The lightest discrete mode corresponds to $n=0$.

On top of the bound states, 
there will be a gapped continuum
in the Higgs decomposition as well,
starting at $\omega^2_\text{KK} > \kappa v_s^2 - \mu^2$ with 
definite parity.
The masses of the continuum states are then given by 
\begin{equation}
 \omega_k^2 = \left( \frac{\tilde \nu_k^2}{2}
 +\frac{\kappa}{\eta}\right) m^2 - \mu^2,
\end{equation}
where $\tilde \nu_k$ is a dimensionless
wavector similarly to $q_k$ in \eqref{eq:sspec}.
The even continuum profiles $\phi^e_k$ are given by
\begin{equation}
 \phi_{k}^e(z) \propto \,\text{Im}
 \left\{
 (1+p+i \tilde \nu_k) P_{p+1}^{-i\tilde \nu_k}(0)
 P_p^{i \tilde \nu_k}(\text{tanh}\,b z)\right\}.
\end{equation}
Far from the domain wall they approach a plane wave form,
\begin{equation}
 \phi_{k}^e(z \rightarrow \infty) \propto \,\text{Im}\left\{ 
 \frac{(1+p+i \tilde \nu_k) P_{p+1}^{-i\tilde \nu_k}(0)}{\Gamma(1-i \tilde \nu_k)}
 e^{i \tilde \nu_k b z}\right\}.
\end{equation}
The allowed values for $\tilde \nu_k >0$ are again dictated
by the boundary conditions in the box $z\in (-L,L)$,
which we shall take to be $\phi^e_k(L)=0$, and they merge into a
continuum for $L \rightarrow \infty$.

For completeness we present the odd solutions $\phi^o_k$ as well,
\begin{equation}
 \phi_{k}^o(z) \propto \,\text{Im}
 \left\{P_p^{i \tilde \nu_k}(0) P_p^{-i \tilde \nu_k}
 (\text{tanh}\,b z)\right\},
\end{equation}
with a similar plane--wave asymptotic behaviour.

Plugging the decomposition above into the 4d action, 
we obtain the quadratic part of the 3d theory,
\begin{equation}\label{eq:quadratic}
 S^{(2)} = \int d^3 x \left[
 \frac{1}{2} (\partial_\mu h_k)^2 
 +\frac{1}{2} (\partial_\mu s_k)^2 - 
 \frac{\omega^2_k}{2} h_k^2
 -\frac{m^2_k}{2} s_k^2\right],
\end{equation}
where the implicit sum over $k$ runs over both the discrete and the 
continuum modes.

The interacting part of the 3d theory can be
obtained from the 4d potential $V$ by performing
the various overlap integrals in the $z$ direction.
This will generate a 3d potential $\tilde V$ with trilinear
and quadrilinear terms in the fields. 
Restricting ourselves for a moment to the lowest--lying
discrete states $h_0$ and $s_0$, we have
\begin{multline}\label{eq:Vdiscrete}
 \tilde V \supset \frac{1}{4} (c_\lambda m) \lambda h_0^4
 + \left( c_{3\eta} m^{3/2} \right) \sqrt{\eta} s_0^3
 + \frac{1}{4} ( c_{4\eta} m) \eta s_0^4 \\
 + \left( c_{3\kappa} m^{3/2}\right) \frac{\kappa}{\sqrt{\eta}}
  s_0 h_0^2 + \frac{1}{2}(c_{4\kappa} m) \kappa h_0^2 s_0^2,
\end{multline}
where the $c$ coefficients are $\lesssim \mathcal{O}(1)$
numbers related to the relevant overlap integrals.
For example, $c_{3\kappa}$ is given by
\begin{equation}\label{eq:g}
 c_{3\kappa} \equiv m^{-3/2} \int_{-L}^L
 \text{d}z \, \sigma_0(z) \sqrt{\eta} S_\text{DW}(z) \phi_0^2(z),
\end{equation}
and depends on the 4d couplings through
the dimensionless combination $\kappa/\eta$.
Similar interaction terms where $h_0$
and $s_0$ are replaced by continuum states will also
appear in $\tilde V$ alongside the ones shown in \eqref{eq:Vdiscrete}.

Since the true vacuum of the 4d theory will have $h=v$
everywhere, we expect only the even Higgs profiles to play a role 
in the phase transition. Similarly, the true vacuum corresponds
to $S=0$, so that only the odd singlet
modes will be excited in order to 
(partly) cancel the unperturbed
domain wall. In addition, we expect
the seeded bubble to have a definite symmetry for $z\rightarrow-z$,
which requires either the odd or the even modes to be identically
zero for each field. In the following we shall therefore 
neglect the odd and 
even states for the Higgs and the singlet, respectively. 
From the point of view of the 3d theory,
we can see that some overlap integrals
for these modes will vanish, so that their vevs
are expected to be set to zero dynamically when minimizing
the potential, leading to the same conclusion. \\

Up to this point the 3d theory under consideration
is equivalent to the 4d theory provided
that all the continuum states are taken into account.
This may be however not possible in practice, and in the following
we shall discuss how the heavy KK states may be 
integrated
out in favor of an effective theory for the lowest--lying
modes.
In particular, we will focus on the theory
obtained by integrating out the two towers
of continuum Higgs and singlet KK modes, thus keeping
only $h_0$ and $s_0$ as 
dynamical\,\footnote{
In fact, the relevant parameter space will be such that $p<1$, so that
there will be a single Higgs bound state with $n=0$ and $\nu=p$. 
If present, additional bound states can nonetheless be included 
straightforwardly.}.
The reason for including $s_0$ in the effective
theory despite its rather large mass
is that we expect its vev to change 
more significantly with respect to the vevs of the continuum states,
as this will be the leading correction to the unperturbed
domain wall.

The two--field effective theory obtained in this way will 
consist of operators made of $h_0$ and $s_0$ up to the sixth
power at the order $\mathcal{O}(1/m^2_\text{KK})$, 
and even in $h_0$.
At the order $\mathcal{O}(1/m^4_\text{KK})$
interactions are generated up to the eighth power together with 
momentum--dependent 
operators\,\footnote{We generically indicate by $m_\text{KK}$ the mass scale
of the continuum modes for both the Higgs and the singlet towers.}. All these new terms should then be 
added to the ones shown in \eqref{eq:Vdiscrete}.

We may estimate the condition for the validity of the effective
theory in terms of the maximum size of the $h_0$ and $s_0$ vevs
by requiring that the induced shift in 
the masses of the continuum modes remains a small correction.
Considering the contribution from $h_0 \neq 0$
to all the singlet KK states, we can then require
\begin{equation}\label{eq:Lambda}
\frac{\Delta m^2}{m^2_\text{KK}}
\equiv \int_0^\infty \frac{\text{d} q}{\sqrt{2} \pi} 
\frac{ c_{4\kappa}(q) m \kappa \,h_0^2}{(q^2/2 + 2)m^2} < 1
\end{equation}
where we have replaced the sum over the states in the box
of size $2L$ with an integral over
$q_k$. The coefficient $c_{4\kappa}(q)$ is the analogue of
$c_{4\kappa}$ in \eqref{eq:Vdiscrete} for the operator
$h_0^2 s_k^2$, in which $s_0$ has now been replaced with states
of the continuum tower.
The integral \eqref{eq:Lambda} 
is dominated by the lightest KK
states due to the $1/q^2$ suppression,
so that we can write 
\begin{equation}\label{eq:Lambda2}
 \frac{\Delta m^2}{m^2_\text{KK}} = \frac{\bar c_{4\kappa} \kappa}
 {2 \sqrt{2} m} h_0^2
\end{equation}
where we have defined $\bar c_{4\kappa} <1/2$ as an
average coupling, resulting in the condition 
\begin{equation}\label{eq:condition}
 h_0^2 < \frac{2 \sqrt{2}}{\bar c_{4\kappa}} \frac{m}{\kappa}.
\end{equation}
A similar result is obtained for $s_0^2$ by estimating
the correction to the Higgs continuum masses.
We thus conclude that the 3d cutoff of the effective
theory is set by the scale 
$\sim \sqrt{m/\kappa}$ up to a order--few factor.
We will use this as an approximate criterium to assess the validity of the
effective theory.
\\

Let us now discuss how temperature corrections
can be taken into account in this picture. 
In general, the terms arising from thermal loops
cannot be fully reabsorbed into a redefinition of the 
tree level couplings in \eqref{eq:V4d}. 
However, in the leading high--temperature approximation
in which only the terms $\propto T^2$ are retained,
these effects can be captured with a simple redefinition
of the scalar mass terms, 
\begin{equation}\label{eq:mumT}
 \mu^2 \rightarrow \mu^2 - c_h T^2, \quad m^2 \rightarrow m^2 - c_s T^2
\end{equation}
where $c_h$ and $c_s$ are given by\,\cite{Espinosa:2011ax}
\begin{equation}
 c_h = \frac{2 m_W^2 + m_Z^2 + m_h^2 + 2 m_t^2}{4 v^2}
 + \frac{\kappa}{12}, \quad c_s = \frac{ 4 \kappa + 3 \eta}{12}.
\end{equation}
The procedure outlined above can be promoted
at finite temperature in a straightforward way by 
replacing the bare masses in \eqref{eq:V4d} with their 
temperature corrected values. 
This also implies that a change in the temperature
can be accounted for by a simple shift in
all the 3d masses in \eqref{eq:quadratic}.
In addition, if the coordinates are measured
in units of $1/m(T)$ the interaction Lagrangian
can actually be kept constant with $T$. In practice this means 
that it is not necessary to perform the full KK reduction 
whenever the temperature is modified.

We note that beyond the high temperature approximation at order $T^2$
the profile of the domain wall
will, in general, no longer respect the functional form in 
\eqref{eq:DW} so that the 3d eigensystem
may be found only numerically, challenging 
the feasibility of our procedure.
On the other hand, the additional terms in the
thermal potential that involve only  the Higgs field 
can in principle be added without spoiling the exact solution
of the eigensystem, with the effect of introducing
more interactions in the 3d potential.
For simplicity we shall however stick to the high temperature
approximation at order $T^2$ in what follows, and neglect in addition 
loop corrections to the potential \eqref{eq:V4d}
as we do not expect these to change the conclusions of 
our analysis.\\

Within these assumptions, we can now study the thermal history 
of the SM plus singlet model from the point of view of its 3d description
and identify the two cases of rolling and tunneling
mentioned in Sec.\,\ref{sec:svsuns}.
These two different scenarios can be distinguished by looking at the sign
of $\omega^2_0$ from \eqref{eq:omegap}
at different temperatures,
\begin{equation}
\label{eq:omega0}
 \omega_0^2(T) = \frac{1}{2} p \,(m^2 - c_s T^2) -\mu^2 + c_h T^2.
\end{equation}
Since by assumption
the electroweak symmetry is still unbroken
at the temperature $T_\text{d}$ (corresponding to the $Z_2$ breaking), 
we will have $c_h T_\text{d}^2-\mu^2 >0$ and thus $\omega^2_0(T_\text{d})$
is positive right after the formation of the domain walls. 
This means that the 3d potential will have a 
global minimum with all the vevs set to zero.

If then $\omega^2_0(T)$ turns negative for $T_\text{d} > T > T_c$,
the domain wall
will become classically unstable and roll towards
a more favorable configuration with 
a non--vanishing Higgs profile inside
the domain wall, in close analogy with 
the bosonic realization of string
superconductivity\,\cite{Witten:1984eb}.
Finding this configuration corresponds to
minimize the 3d potential $\tilde V$ in terms of the vevs
of the discrete and continuum modes. 
When the temperature drops below
the critical temperature, the true vacuum of the 
theory is the electroweak one where $S=0$,
and we then expect this 3d minimum to turn 
unstable and the domain wall to dissociate.
This case will be discussed in more detail 
in Sec.\,\ref{sec:rolling}.

The other possibility is that the Higgs eigenvalue $\omega^2_0(T)$
may turn negative only at $T<T_c$,
or actually stay positive even at zero temperature.
Considering the effective theory where the KK towers
are integrated out,
this means that the 
origin in the $(h_0,s_0)$ plane,
where $\tilde V = 0$, is a local minimum. 
The actual location of the global minimum is instead beyond
the capability of the effective theory: this is because
the configuration with $h(x,z)=v$ and $S(x,z)=0$ 
corresponds in the 3d picture to the whole tower of KK states taking non--negligible
vevs. 
Nonetheless, if the 3d potential turns negative
for field values close to the origin,
the seeded false vacuum decay can be studied as a
standard
tunneling process
\begin{equation}\label{eq:3ddecay}
 (0,0)_\text{3d} \rightarrow (\langle h_0 \rangle,
 \langle s_0 \rangle)_\text{3d},
\end{equation}
provided that the release
point is within the validity of the effective theory
\footnote{For
consistency, we also need to require that the size of the bubble
nucleated inside the domain wall is smaller than the correlation
length of the network, such that the wall can be considered straight
as assumed implicitely in \eqref{eq:DW}. This is however a mild
constraint given that the characteristic scale
of the network is horizon--size in the scaling regime, and we shall
always assume that this condition is fulfilled.}.
Having calculated the release point in terms of $h_0$ and $s_0$,
the actual profile of the 4d seeded bubble can be reconstructed
through the
KK decomposition in \eqref{eq:ansatz}, 
where the vevs of the continuum states are fixed by their 
equations of motion that we have used to integrate them out.

The tunneling process \eqref{eq:3ddecay} can
be solved with usual techniques. In the high temperature
limit, it consists of finding a time--independent solution
with $O(2)$ symmetry to the 
system of ordinary differential equations 
\begin{equation}\label{eq:bounceT}
 \phi^{\prime \prime}_i + \frac{1}{r} \phi^\prime_i
 = \frac{\partial \tilde V}{\partial \phi_i}, \quad \phi_i=h_0,s_0
\end{equation}
which can be solved e.g. with \texttt{CosmoTransitions}\,\cite{Wainwright:2011kj},
with $r^2=x^2+y^2$ being the radius in the domain wall plane,
and boundary conditions $\phi_i(\infty)=0$ and $\phi^\prime_i(0)=0$
ensuring that the unperturbed domain wall is recovered at $r = \infty$.
The bounce action corresponding to this solution will be indicated
by $S_2/T$, where the suffix refers to the $O(2)$ symmetry,
and will be studied in detail in Sec.\,\ref{sec:tunneling}
and \,\ref{sec:bubble}.

The same strategy may be applied
to search for solutions with a spherical symmetry $O(3)$ 
involving the radial variable
on the domain wall and the euclidean time $\tau$,
\begin{equation}\label{eq:bounceO3}
 \phi^{\prime \prime}_i + \frac{2}{\rho} \phi^\prime_i
 = \frac{\partial \tilde V}{\partial \phi_i},
 \quad \phi_i=h_0,s_0
\end{equation}
where $\rho^2=r^2+\tau^2$, which can be solved with help of 
\texttt{CosmoTransitions} and \texttt{FindBounce}\,\cite{Guada:2020xnz}.
The corresponding $\tilde S_3$ action, where the suffix
refers to the $O(3)$ symmetry, is expected to become
smaller than $S_2/T$ only at very low temperatures and will play no role 
in our analysis, as domain walls are anyway destroyed
at the time of the electroweak phase transition. 
Let us however note that $\tilde S_3$ can
be relevant in different scenarios where the defects survive until today
and may then affect the SM life time.

Let us now briefly discuss how the boundary conditions
for the bounce along $z$ are satisfied in the approach
discussed above.
These conditions are such that the false vacuum $(0,\pm v_s)$ needs
to be recovered
at $z\rightarrow \pm \infty$. 
As far as the discrete modes $h_0$ and $s_0$ are concerned, 
this is automatically the case because the localized profiles 
$\phi_0(z)$ and $\sigma_0(z)$ vanish at $|z|=\infty$.
On the other hand, scattering states will oscillate
with a certain amplitude far away from the domain wall,
and only an infinite superposition may lead to the 
exact boundary condition. However, already the 
individual amplitude of these oscillations is expected
to be small given that they are suppressed by 
the KK scale, so that no additional condition 
needs to be imposed besides the validity of the effective theory.

Finally, let us mention that there is 
one caveat to this procedure when applied
to the effective theory at $\mathcal{O}(1/m^4_\text{KK})$
due to the occurrence of momentum--dependent operators. These would 
in fact modify the equations of motion \eqref{eq:bounceT}
and \eqref{eq:bounceO3}
beyond a simple change in the shape of the potential $\tilde V$,
which would then require a non--trivial
modification of the solving procedure.
In our analysis we will neglect the derivative
operators when evaluating $S_2/T$ at  $\mathcal{O}(1/m^4_\text{KK})$
and leave this improvement for future work.

\subsection{Thin wall approximation}
\label{sec:TW}

In this section we shall discuss an alternative approach, namely
the thin wall (TW) approximation, 
to calculate the inhomogeneous tunneling
rate induced by the domain walls.
While the TW approximation has limited validity for what concerns the temperature range,
it has the virtue of providing further insights in the physics of the seeded tunneling, and,
as we shall see in Sec.\,\ref{sec:results}, it can nicely
complement the KK method described above.

In order to evaluate the transition probability at finite temperature,
we need to estimate the change in energy between the bounce
configuration and the initial configuration.
As a first approximation we can assume that the bounce respects
an $O(3)$ symmetry and that it is described by a spherical bubble of radius $R$
centered on the domain wall.
The difference in energy between the bubble configuration and the false vacuum
(corresponding to the unperturbed domain wall)
is
\be
\label{energy_bounce}
E(R)=4 \pi  R^2 \sigma_\text{B} -\frac{4}{3} \pi R^3 \Delta V - 
\pi R^2 \sigma_\text{DW}.
\ee
Here $\sigma_\text{B}$ is the bubble tension and $\Delta V $ is the (positive) potential difference between the false and the true vacuum. 
The domain wall tension is 
\be
\sigma_\text{DW} = \frac{4}{3} \sqrt{\frac{\eta}{2}} v_s^3,
\ee
where $v_s$ here denotes the temperature dependent vev of $S$.
The first two terms in \eqref{energy_bounce} are the usual surface term and 
volume term of a first order phase transition in 
homogeneous space\,\cite{Coleman:1977py,Linde:1981zj,Anderson:1991zb}.
Assuming that the presence of the domain wall does not affect significantly the shape and tension of the bubble,
the tension $\sigma_\text{B}$ can then be estimated with standard techniques developed for first order phase transitions. 
For the case of a model with two scalar fields we illustrate a procedure to derive it in 
Appendix\,\ref{app:bubble_tension}.
The last term in \eqref{energy_bounce} is due to the presence of the domain wall, and characterizes the energy difference 
between the initial configuration
with an extended domain wall, and the bounce configuration where the domain wall has a hole of radius $R$. 
The critical radius for the bubble to be created thermally can be obtained by extremizing the energy with respect to $R$, 
finding $R_c= \frac{2}{\Delta V }\left(\sigma_\text{B} -\sigma_\text{DW}/4 \right)$.

\begin{figure}
\centering
 \includegraphics[scale=0.7]{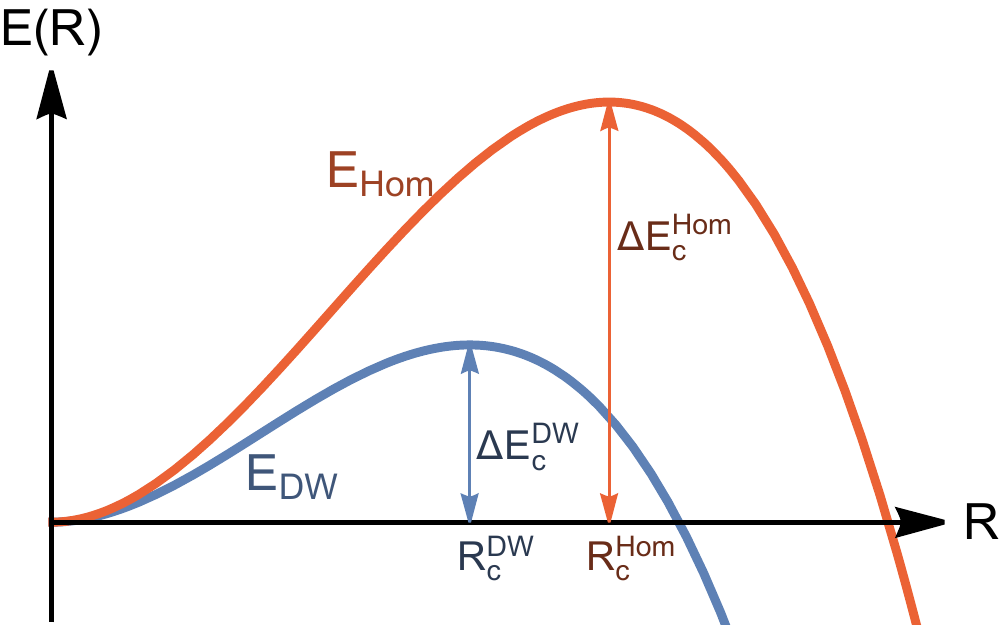}
 \caption{
Energy of the bubble as a function of the radius. In red the standard case of homogeneous phase transition. 
In blue the case of a phase transition seeded by domain walls, where the bubble is encircling a disk of radius $R$ on the domain wall. 
For the two cases, $R_c^{\text{Hom}}$ and $R_c^{\text{DW}}$ denote the critical bubble radius, while $\Delta E_c^{\text{Hom}}$ and
$\Delta E_c^{\text{DW}}$ the energy necessary for the bubble to form.
The inhomogeneous process results energetically favorable.
\label{fig:cartoon}
 } 
\end{figure}

In Fig.\,\ref{fig:cartoon} we display a representation of the energy of the nucleated bubble as a function of the radius $R$, both
for the case without the domain wall corresponding to the homogeneous transition, and for the case with the domain walls.
As it is clear also from Eq.\,\eqref{energy_bounce}, 
the presence of the domain wall makes the energy necessary to reach the critical radius smaller than in the case of the homogeneous transition.
Note that this does not guarantee a priori that the seeded phase transition will dominate over the homogeneous one, since the nucleation condition 
is different, see Eqs.\,\eqref{eq:nucle_homo} and \eqref{eq:nucDW}.

Given the expression for the energy \eqref{energy_bounce}, 
the rate for the thermal transition is $\propto \text{exp}(-E(R_c)/T)$, and hence we can estimate the thermal bounce action as
\be
\label{bounce_TW_O3}
\left(\frac{S_3}{T}\right)_{\text{TW}} \simeq 
\frac{16 \pi \left(\sigma_\text{B} -\sigma_\text{DW}/4 \right)^3}
{3 T \Delta V^2},
\ee
where here the subscript indicates that we have considered an $O(3)$ spherical bubble geometry.
Note that in the limit $\sigma_\text{DW} \to 0$ this expression reduces to the standard $O(3)$ bounce action 
for bubble nucleation for a
homogeneous phase transition
at finite temperature \cite{Linde:1981zj}.
The net effect of the presence of the domain wall, in comparison with the homogeneous tunneling, is to reduce the bubble tension to an effective 
tension $\sigma_\text{B} -\sigma_\text{DW}/4$.

Let us now discuss how the previous estimate of the thin wall bounce action can be improved
by taking into account 
that the symmetry of the bubble is actually not $O(3)$ but only $O(2)$.
Indeed, when the domain wall tension is large and comparable to the bubble tension, the bubble of true vacuum has a lower energy 
if it gets flatter around the domain wall.

One may then consider
different geometrical shapes for the bubble. 
A simple ansatz respecting a $O(2)$ symmetry is a spheroid centered on the domain wall 
with the two radii on the domain wall 
plane taken to be equal.
In this case the energy difference between the bounce and the initial configuration is
\be
\label{energy_bubble_O2}
E(R_1,R_2) = \Sigma(R_1,R_2) \sigma_\text{B} -\frac{4}{3} \pi R_1^2 R_2\Delta V - \pi R_1^2 \sigma_\text{DW} 
\ee
where $R_1$ is the size of the two radii in the plane of the domain wall, and $R_2$ 
is the radius perpendicular to the domain wall plane.
$\Sigma(R_1,R_2)$ denotes the spheroid surface.
Given the tensions and $\Delta V$, this energy should be extremized numerically
for $R_1$ and $R_2$ to find the critical radii $R_1^c$ and $R_2^c$.
The thermal bounce action is then estimated in this case as 
$(S_2/T)_{\text{TW}} \simeq E(R_1^c,R_2^c)/T $.
 Note that in principle, the bubble tension in \eqref{energy_bubble_O2} is also 
 a function of the radii $R_1$ and $R_2$, since it can change as a function of 
 the curvature on the spheroid.
In the following, we will instead approximate $\sigma_B$ in \eqref{energy_bubble_O2} 
with the tension computed assuming $O(3)$ symmetry, 
which is an underestimate of the $O(2)$ tension.

Let us finally note that when the tension of the domain wall, $\sigma_\text{DW}$, becomes of 
the order of the bubble tension, $\sigma_\text{B}$, the critical radius in Fig.\,\ref{fig:cartoon}
goes to zero and so does the bounce action, signalling that the phase transition is not first order.
This behaviour actually corresponds to the same classical instability described in the previous sections, 
and it occurs at $\sigma_\text{B} \sim \sigma_\text{DW}/4$ and 
$\sigma_\text{B} \sim \sigma_\text{DW}/2$ for the $O(3)$ and $O(2)$ bounces, respectively.
However, this typically happens at temperatures significantly lower than $T_c$
where the thin wall approximation is not precise, and in
the following we shall
rely only on the condition discussed around Eq.\,\eqref{eq:omega0} to identify this behaviour.

\section{Results}
\label{sec:results}

The Higgs--singlet model depends on three independent parameters 
($m_S,\kappa,\eta$) according to 
the potential \eqref{eq:V4d}, where $m_S$ is the singlet mass
in the electroweak vacuum as given in \eqref{eq:ms}.
As mentioned in Sec.\,\ref{sec:setup},
we will here focus on the region of parameter space where a 
two--step phase transition occurs, so that the domain walls
formed in the first step can act as seeds for the second one.

\begin{figure}
\centering
 \includegraphics[scale=0.38]{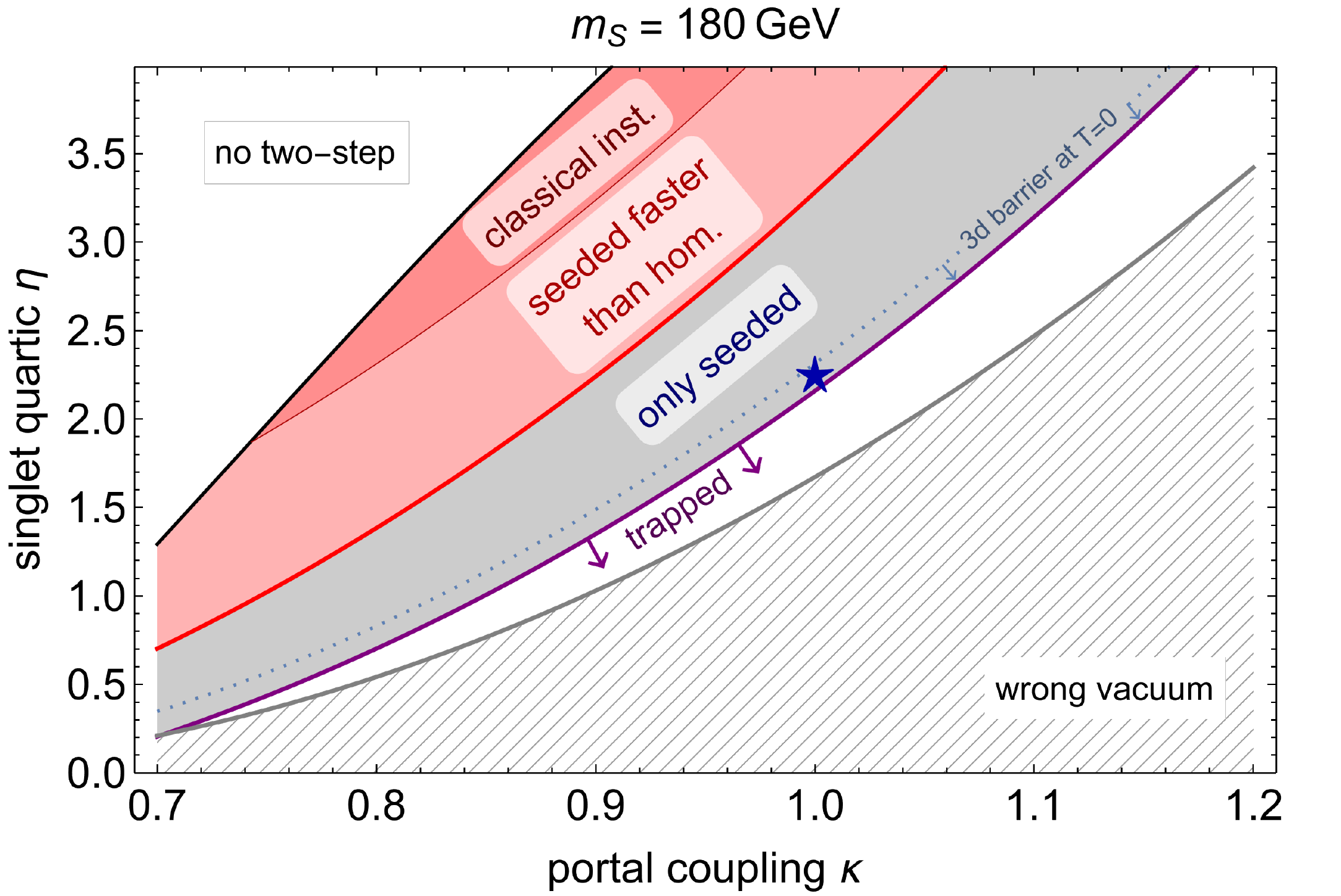}
 \caption{Scan in the $(\kappa,\eta)$ parameter
 space for $m_S = 180$ GeV.
 The upper--left corner is excluded from our analysis
 as the phase transition does not follow
 the two--step path described in Sec.\,\ref{sec:setup}.
 The bottom--right corner is also excluded as the electroweak
 minimum is not the true vacuum at zero temperature.
 The regions shaded in red indicate where a first order phase
 transition is expected due to homogeneous tunneling.
 Our analysis shows that the seeded false vacuum decay
 is instead the dominant process for completing the phase transition:
 in the dark red region this is because of a classical
 instability of the domain walls (rolling), whereas in the light
 red region this is due to the seeded nucleation being
 faster than the homogeneous one. In the blue region 
 the homogeneous bubbles would actually fail to nucleate and the only
 relevant process there is seeded nucleation. Below the dotted blue line
 the barrier for the seeded tunneling persists also at zero temperature,
 and the fields may be trapped if the nucleation rate is too small.
 This is the case for the points below the purple
 line which are then not viable.
 The benchmark point close to the no--nucleation line 
 indicated by the blue star
 will be discussed in detail in Sec.\,\ref{sec:bubble}.}
 \label{fig:scan180}
\end{figure}

The general outcome of our study
is that the seeded vacuum decay is always found to be
the fastest process compared to the homogeneous tunneling.
In addition, regions of the parameter
space where the fields would stay trapped if domain walls
were neglected
become now viable due to the seeded tunneling.
An overview of our results is shown in Fig.\,\ref{fig:scan180}
where we have performed a scan over the parameter space in terms
of the portal coupling $\kappa$ and the singlet quartic $\eta$,
fixing the singlet mass to $m_S = 180$ GeV as a representative
value.

The red--shaded regions in Fig.\,\ref{fig:scan180}
are those where the homogeneous phase transition
is expected to be first order. The corresponding
tunneling rate is sufficiently fast until the red solid line
is reached, 
below which bubbles fail to nucleate.
When taking into account the presence of the domain walls, however, 
the phase transition is found to proceed
always via seeded false vacuum decay
in all the parameter space
according to either of the mechanisms
of Sec.\,\ref{sec:svsuns}. 
In the dark red region
this is due to the
domain walls developing a classical instability
above the critical temperature and then dissociating
as soon as the temperature drops below $T_c$
(see Sec.\,\ref{sec:rolling}).
In the light red region a classical instability
is actually never developed, and the transition is
completed via seeded bubble nucleation
(see Sec.\,\ref{sec:tunneling}).

Within the blue region in Fig.\,\ref{fig:scan180}, the phase
transition can instead only proceed via seeded vacuum decay. 
The benchmark point indicated
by the blue star in Fig.\,\ref{fig:scan180} will be 
studied in detail in Sec.\,\ref{sec:bubble}
where we estimate the nucleation temperature as well
as the shape of the nucleated bubble.
The seeded tunnelling stops being cosmologically
fast below the purple line where the fields
will remain trapped in the false vacuum.

\subsection{The rolling}
\label{sec:rolling}

We here discuss in more detail the case in which the 
domain walls become unstable already above the critical
temperature.
Let us assume that the first instability occurs at $T_r > T_c$
and corresponds to $\omega_0^2(T_r)$ turning negative. 
This signals 
that the linearized equation of motion for the 4d Higgs field 
in the domain wall background supports 
at least one solution in which perturbations grow 
exponentially in time.
Since we are still above $T_c$, 
this will eventually result in a new (stable)
domain wall configuration with a non--vanishing
Higgs profile.
When the temperature drops below $T_c$,
these configuration will eventually dissociate completing the phase transition.

We shall first consider classical instability in a toy benchmark
for which the new stable configuration
above $T_c$ can be obtained exactly,
and later on discuss the relevance for 
the electroweak phase transition.
This will allow us to validate our method
based on the KK decomposition presented
in Sec.\,\ref{sec:formalism} by comparing it against
these exact results.

Setting $\eta=1$ and $m =1$ (which is 
possible thanks to field redefinitions and rescaling
of the space time coordinates)
the potential \eqref{eq:V4d} will have a global minimum
at $S=\pm 1$ for $\lambda > \mu^4$ allowing for the 
domain wall solution \eqref{eq:DW}
(the temperature dependence is here left implicit). 
Focussing on the case where the remaining parameters are such
that $\omega_0^2<0$, the new stable configuration
can be obtained exactly
for very special values of $\lambda$, 
see e.g.\,\cite{Rajaraman:1982is},
namely $\lambda = \kappa(\kappa-2\mu^2)/
(2\kappa-2\mu^2-1)$. 
The vacuum structure of this toy benchmark is then actually
equivalent to have a Higgs instability above $T_c$, with the
advantage that the stable profiles
are known:
\begin{align}
\label{eq:exactS}
&S(z)= \,\text{tanh}(z/\delta), \\
\label{eq:exacth}
&h(z) =\sqrt{1/\kappa + 2\mu^2/\kappa-2}\,
 \text{sech}(z/\delta),
\end{align}
where $\delta = (\kappa-\mu^2)^{1/2}$. 

We can now try to approximate this exact result
in our 3d theory, where
the problem of finding the stable 
configurations \eqref{eq:exactS}
and \eqref{eq:exacth}
is equivalent to searching
for the minimum of the potential
$\tilde V$. 
Discrete and scattering
states will then take a vev in the new vacuum,
thus modifying the overall profiles along $z$
according to the superposition in \eqref{eq:ansatz}.

In practice there are three ways to minimize the 3d potential: 
\textit{i)} Include a (large enough) number
of scattering states together with the discrete
modes and minimize a multi field potential.
This way is the most reliable but it may become 
unpractical for a large number of fields.
\textit{ii)} Include a (large enough) number of scattering states
and integrate them out at tree level in favour
of an effective potential for $h_0$ and $s_0$ only,
according to the discussion in Sec.\,\ref{sec:KK}.
\textit{iii)} Neglect the scattering states altogether
and minimize the quartic potential for $s_0$ and $h_0$
in \eqref{eq:Vdiscrete}.
This corresponds to the lowest level of accuracy.

\begin{figure}
\centering
 \includegraphics[scale=0.48]{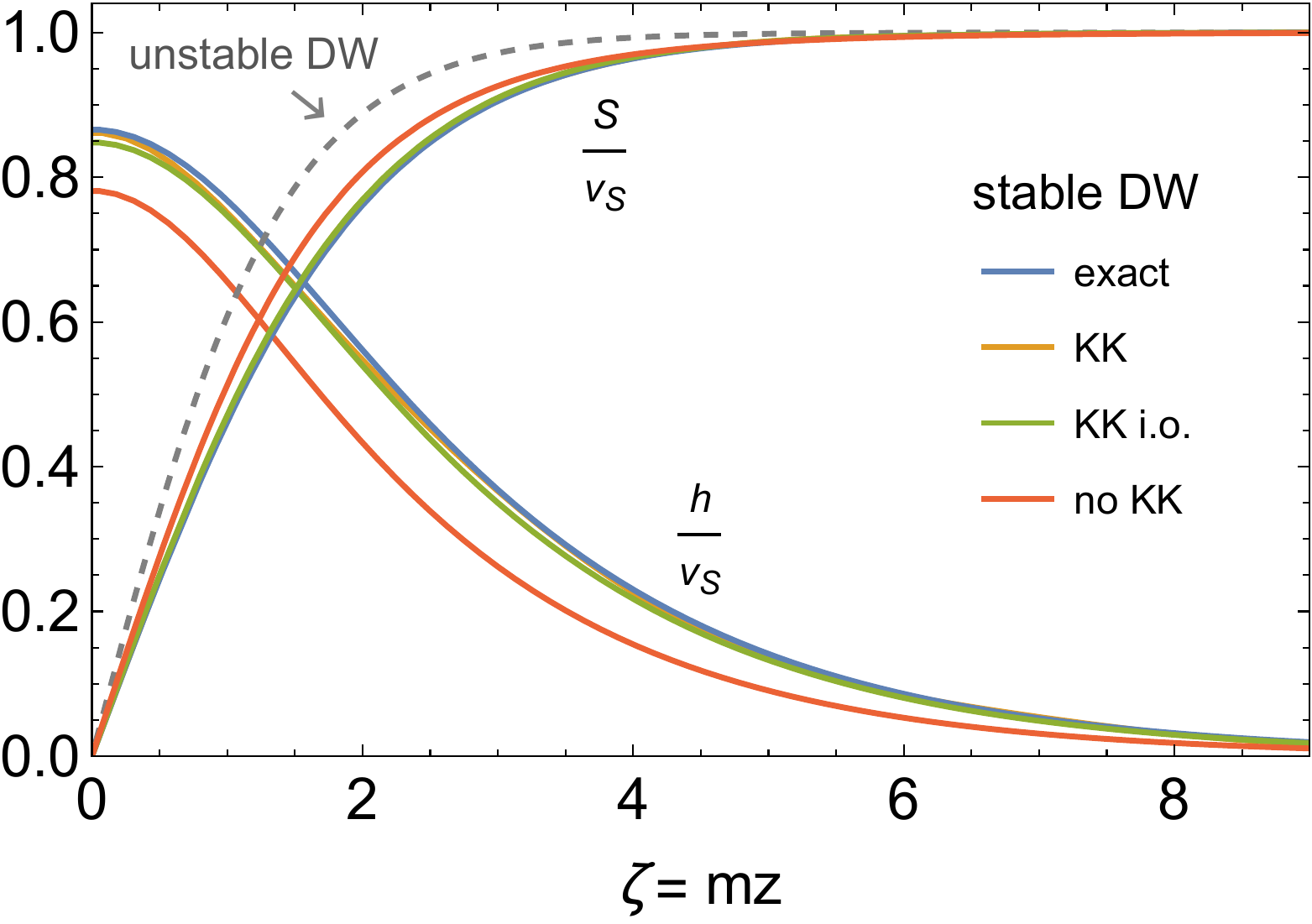}
 \caption{The new stable configuration 
 approached by the system following an instability
 of the unperturbed domain wall above the critical
 temperature in a toy benchmark of the Higgs--singlet model. 
 The $h$ and $S$ profiles are respectively
 even and odd for $z\rightarrow -z$. 
 The dashed gray line shows the unperturbed domain wall
 which turns unstable when $\omega^2_0 <0$.
 Blue lines show the exact profiles of the stable configuration.
 The remaining colorful lines approximate the exact solutions
 according to the 3d methods discussed
 in the text: \textit{i)} minimizing the multi field potential
 (orange), \textit{ii)} integrating out the KK fields at tree level
 and minimize the effective two--field potential for the discrete states (green), 
 \textit{iii)} neglect altogether the KK states and minimize
 the two--field potential for the discrete states (red).
 Once the minimum in terms of the 3d fields is found in the various
 approximation schemes, the overall profile
 along $z$ can be reconstructed via \eqref{eq:ansatz}.
 The model parameters were taken to be $m=1$,
 $\eta=1$, $\mu^2=5/12$ and $\kappa=2/3$ (with these choices
 one also has $v_s = 1$).}
 \label{fig:raja}
\end{figure}

We compare these different options
in Fig.\,\ref{fig:raja} for a representative choice
of $\mu^2$ and $\kappa$.
The exact solutions
\eqref{eq:exactS} and 
\eqref{eq:exacth} are shown by the blue lines.
The results employing
\textit{i)} are shown in orange, 
where we have included KK states up to masses
$\sim 10 |\omega_0|$ for the two towers,
and taking the size of the box to be $2L \simeq 50 \, m^{-1}$.
The green line shows the solution
integrating out KK states according to \textit{ii)}
up to masses $\sim 15 |\omega_0|$
with $\mathcal{O}(1/m_\text{KK}^4)$ accuracy, using the 
same value for the size of the box.
Finally,
the red line is obtained by neglecting altogether the KK 
contribution to the potential according
to \textit{iii)}, and is based
on the terms given in \eqref{eq:Vdiscrete}.

As we can see, options \textit{i)} and \textit{ii)} 
provide a quantitative
agreement with the exact result, and improve
on the simplest approach \textit{iii)}. 
The location of the global minimum in terms of
$h_0$ turns out to be $h_0 \simeq 1.7$,
and the ratio in \eqref{eq:Lambda2} evaluates to 
$\sim 0.3$ confirming the reliability of the expansion.

After having validated our 3d methods by reproducing
these non--trivial
results, let us now move to study more phenomenologically
interesting benchmarks in the singlet--extended SM,
where the unperturbed domain wall
can develop an instability above $T_c$ in
some part of the parameter space.

We find that this is in fact the case
in the region shaded in dark red 
in Fig.\,\ref{fig:scan180} for the choice $m_S=180$ GeV,
(whereas this is never realized in our other scan
in Fig.\,\ref{fig:scan250} with $m_S = 250$ GeV).
For these points we find that
the rolling temperature $T_r$ is only slightly above 
the critical temperature,
so that a stable domain wall with a Higgs core may last only
shortly. When the temperature drops below $T_c$,
no global minimum is found in the region where the effective
theory is valid, in agreement with the 
intuition that domain walls
with a non--vanishing Higgs profile should start
dissociating when the true vacuum is the one
with $S=0$.

The seeded vacuum decay in this case
may resemble a very weakly first--order 
electroweak phase transition occurring
at $T_c$,
with the notable difference that the process
is not controlled by the bubble nucleation rate,
but rather by the number of domain walls per hubble patch
and their velocity\,\cite{PhysRevD.30.272}.
This means that the time for completing the
phase transition may differ significantly when compared
to this first--order analogue.
Possible phenomenological implications are discussed
in more detail in Sec.\,\ref{sec:discussion}.

\subsection{The tunneling}
\label{sec:tunneling}

Let us now move to investigate the other possibility in which 
the mass of the discrete Higgs mode $\omega^2_0(T)$
is positive for a range of temperatures below $T_c$,
implying that the seeded phase transition will proceed
via bubbles nucleated inside the domain walls.

\begin{figure}
\centering
 \includegraphics[scale=0.4]{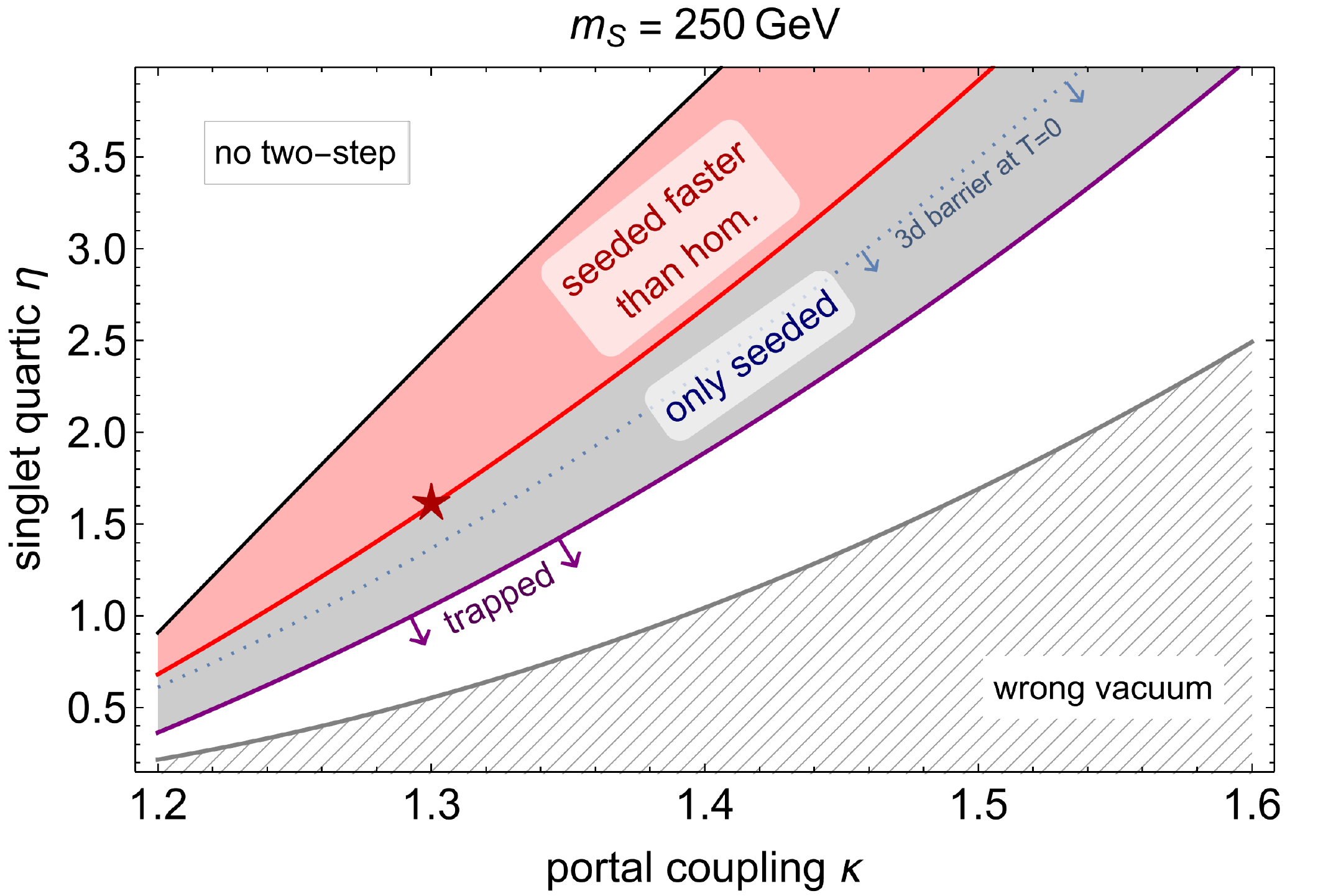}
 \caption{Scan in the $(\kappa,\eta)$ parameter
 space for $m_S = 250$ GeV.
 The meaning of the various labels is the same as in 
 Fig.\,\ref{fig:scan180}. In the light red region a two--step 
 first order phase transition is expected due to homogeneous
 tunneling. Seeded nucleation is however faster and it will be the leading
 mechanism for completing the phase transition.
 In the blue region homogeneous bubbles fail to nucleate but the 
 transition still completes due to the seeded vacuum decay.
 Below the purple line even the seeded nucleation becomes
 cosmologically slow and the fields are trapped in the false vacuum.
 The point marked by the red star is studied in detail 
 in Fig.\,\ref{fig:ben250}.}
 \label{fig:scan250}
\end{figure}

As mentioned at the beginning of this section,
this process is the relevant one in the parameter
space shaded in light red in Fig.\,\ref{fig:scan180}
with $m_S=180$ GeV.
The same behaviour persists also when changing
the mass of the singlet scalar to $m_S = 250$ GeV,
as shown in Fig.\,\ref{fig:scan250}
(with the only qualitative difference that the region of classical
instability is no longer found).

\begin{figure}
\centering
 \includegraphics[scale=0.447]{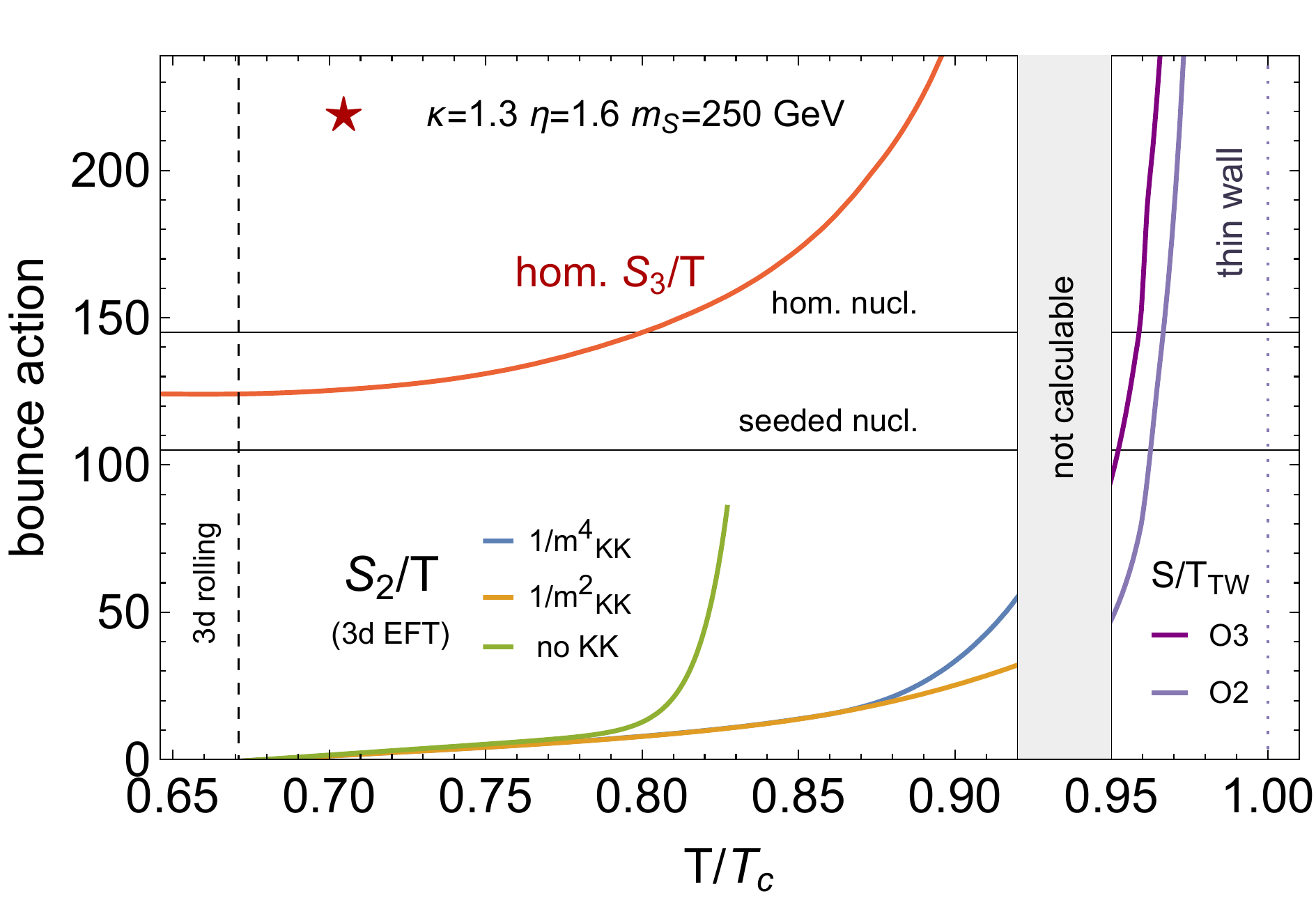}
 \caption{Benchmark point for $m_S = 250$ GeV
 with $\kappa = 1.3$ and $\eta = 1.6$.
 The gray horizontal lines indicate the thresholds for a cosmologically
 fast tunneling at the electroweak scale for a homogeneous
 phase transition and for a transition seeded by domain walls, 
 as discussed in Sec.\,\ref{sec:svsuns}.
 The dashed vertical line indicates the temperature at which
 the (thermal) barrier for the seeded tunneling vanishes
 so that the corresponding bounce action approaches zero.
 The red line refers to the $S_3/T$ action for the homogeneous
 tunneling. The inhomogeneous action $S_2/T$ is shown according
 to the various approximations discussed in the text,
 namely neglecting the continumm KK states
 (green), in the effective theory at $\mathcal{O}(1/m^2_\text{KK})$
 (orange) and $\mathcal{O}(1/m^4_\text{KK})$ (blue) accuracy,
 and in the thin wall approximation,
 which is reliable close to the critical temperature, 
 with a $O(3)$ (purple) and $O(2)$ (light purple) ansatz for the bubble. 
 By this latter method we estimate the
 nucleation temperature to be $\sim 0.96 \, T_c$ in this benchmark.
 The gray--shaded area corresponds to the temperature range in which 
 none of the various methods implemented here can be applied reliably.
 }
 \label{fig:ben250}
\end{figure}

In order to assess whether the phase transition will be completed
via homogeneous or seeded tunneling, we have compared
the homogeneous action $S_3/T$
with the inhomogeneous
action $S_2/T$,
taking into account the different nucleation conditions 
derived in Sec.\,\ref{sec:svsuns}.

The bounce action $S_3/T$ can be evaluated
as a function of the temperature with the help of
\texttt{CosmoTransitions} and \texttt{FindBounce}
from the temperature--dependent 4d potential.
As mentioned in Sec.\,\ref{sec:formalism},
a full--fledged implementation of the thermal
corrections is challenging for the formalism
based on dimensional reduction. In order to make
a consistent comparison
we shall then use the high temperature approximation
also for $S_3/T$.

The action $S_2/T$ can be computed in the 3d theory
following three levels of approximations 
similarly
to what discussed in the previous subsection.
The difference here is that the option \textit{i)}
that keeps the KK states in the theory
becomes quickly
unpractical as this would require to solve 
a multi--field bounce.
For this reason, we will only study
the two--field problem for the discrete modes
$h_0$ and $s_0$ with different level of accuracy,
namely in the effective theory where the continuum
KK states are integrated out at
$\mathcal{O}(1/m_\text{KK}^4)$ and $\mathcal{O}(1/m_\text{KK}^2)$,
and neglecting the KK states altogether.
Once the effective potential is obtained with these different
approximations, we can find the bounce solution
with the help of \texttt{CosmoTransitions}
since the $O(2)$ bounce requires to set 
$d=2$ for the spacetime dimension.

A comparison between the various tunneling rates for the benchmark
indicated by the red star in Fig.\,\ref{fig:scan250} with 
$m_S = 250$ GeV, $\kappa=1.3$ and $\eta = 1.6$,
is shown in Fig.\ref{fig:ben250}. For this benchmark
the critical temperature is $T_c \simeq 110$ GeV,
whereas the spontaneous $Z_2$ breaking happens
at $T_\text{d} \simeq 140$ GeV.
The 4d theory is such that the barrier between
the false $(0,v_s)$ and true $(v,0)$ vacuum
is non--vanishing also at zero temperature. On the other hand,
the $h_0$ mass in the 3d theory turns negative at 
the would--be rolling temperature $T_r \simeq 0.67
\, T_c$,
meaning that the barrier for the seeded tunneling is only thermal. 

The homogeneous bounce action is plotted in red
and the nucleation condition
is satisfied at $T \sim 0.8 \,T_c$.
The action $S_2/T$ for the seeded tunneling is shown
according to the various approximations
at the order $\mathcal{O}(1/m^4_\text{KK})$ (blue),
$\mathcal{O}(1/m^2_\text{KK})$ (orange), and neglecting
the continuum states (green). 
At low temperatures close to the rolling at $T_r$,
the barrier is small and the release point is very close to the origin. 
The effective operators arising from the exchange of the KK
states are thus suppressed and all the approximation schemes
give the same result. 
At higher temperatures the effect of the continuum
states is more and more important, and around $T/T_c \sim 0.82$
it becomes crucial in making the origin 
$(0,0)_\text{3d}$ metastable. For this reason the 
prediction obtained considering only the discrete states 
starts departing from the other two.

The effective theories at $\mathcal{O}(1/m^2_\text{KK})$
and $\mathcal{O}(1/m^4_\text{KK})$ still agree with each other
for $T \lesssim 0.9 \, T_c$. For higher temperatures 
the release point starts to leave the region of validity,
with \eqref{eq:Lambda} being $\gtrsim \mathcal{O}(1)$,
and calculability is lost. This 
corresponds to the gray--shaded region in Fig.\,\ref{fig:ben250}.
In fact, very close to $T_c$ the energy difference between
the false and true vacuum is small. This means that
a larger bubble is needed for the energy gain
to compensate its tension. This bubble will then become
more and more spherical, thus probing more and more the direction
orthogonal to the domain wall. From the point of view of our 3d theory,
this requires a large number of KK continuum states to be excited,
thus explaining the loss of validity of the effective theory.

However, in this very limit the bubble wall becomes thin,
so that we can apply the approximation discussed in Sec.\,\ref{sec:TW}.
The prediction according to 
the $O(3)$ and $O(2)$ bounce actions in the thin wall limit are
shown in Fig.\,\ref{fig:ben250} by the purple and light purple
line, respectively, until $T = 0.95\, T_c$,
below which the thin wall approximation becomes less accurate.
This information can nonetheless be used to estimate
the nucleation temperature to be around 
$T \sim 0.96\,T_c$ in this benchmark,
and to complement the effective
theory calculation of $S_2/T$ in the region close 
to the critical temperature.
For illustration, in Appendix\,\ref{app:estimate} 
we employ the thin wall approximation to estimate the nucleation temperature 
also for other points in the parameter space explored in 
Fig.\,\ref{fig:scan250}.

We can then summarize our findings as follows.
First of all, it is clear that in this benchmark
point the seeded tunnelling is faster
than the homogeneous one, since at the time of the 
would--be homogeneous nucleation the seeded bounce action is well below
its nucleation condition. The light red regions
in Fig.\,\ref{fig:scan180} and Fig.\,\ref{fig:scan250} are in fact
identified in this way.

Secondly, we have encountered a rather
generic feature of the 3d effective theory, namely the fact 
that it works the best at lower temperatures where the bubble
is thick, and that it will necessarily break down close to the critical
temperature. This shows an interesting complementarity with the
thin wall approximation, which is instead most reliable in the opposite
regime close to $T_c$. In general, a gap of calculability is expected 
for intermediate temperatures, as indicated by the gray zone in 
Fig.\,\ref{fig:ben250}. 

\subsection{The bubble}
\label{sec:bubble}

As discussed in the previous subsection, the closer the seeded
nucleation is to the critical temperature the more
we can expect the bubbles to approach a spherical symmetry.
Away from this limit, instead, the presence of the catalyzing 
domain wall
will generally favor a spheroidal shape 
in which the size of the critical bubble
along the domain wall plane is bigger than 
the one in the orthogonal $z$ direction. 

Let us here investigate the properties of the 
seeded bubbles in detail within the 3d formalism for the benchmark
point indicated by the blue star in Fig.\,\ref{fig:scan180}
with $m_S = 180$ GeV, $\kappa=1$ and $\eta=2.2$.
The critical temperature is $T_c \simeq 60$ GeV,
whereas $T_\text{d} \simeq 180$ GeV.
In this region of parameter space the homogeneous nucleation is 
very slow and the transition can be completed only via seeded tunneling.
This benchmark is also rather close to the no--nucleation condition,
so that we can expect the transition to take place significantly
below the critical temperature, where the
thin wall approximation can be hardly applied.
In addition, the 3d barrier here persists also at 
$T=0$ so that the bounce action $S_2/T$ will asymptotically
grow at low temperatures.

\begin{figure}
\centering
 \includegraphics[scale=0.39]{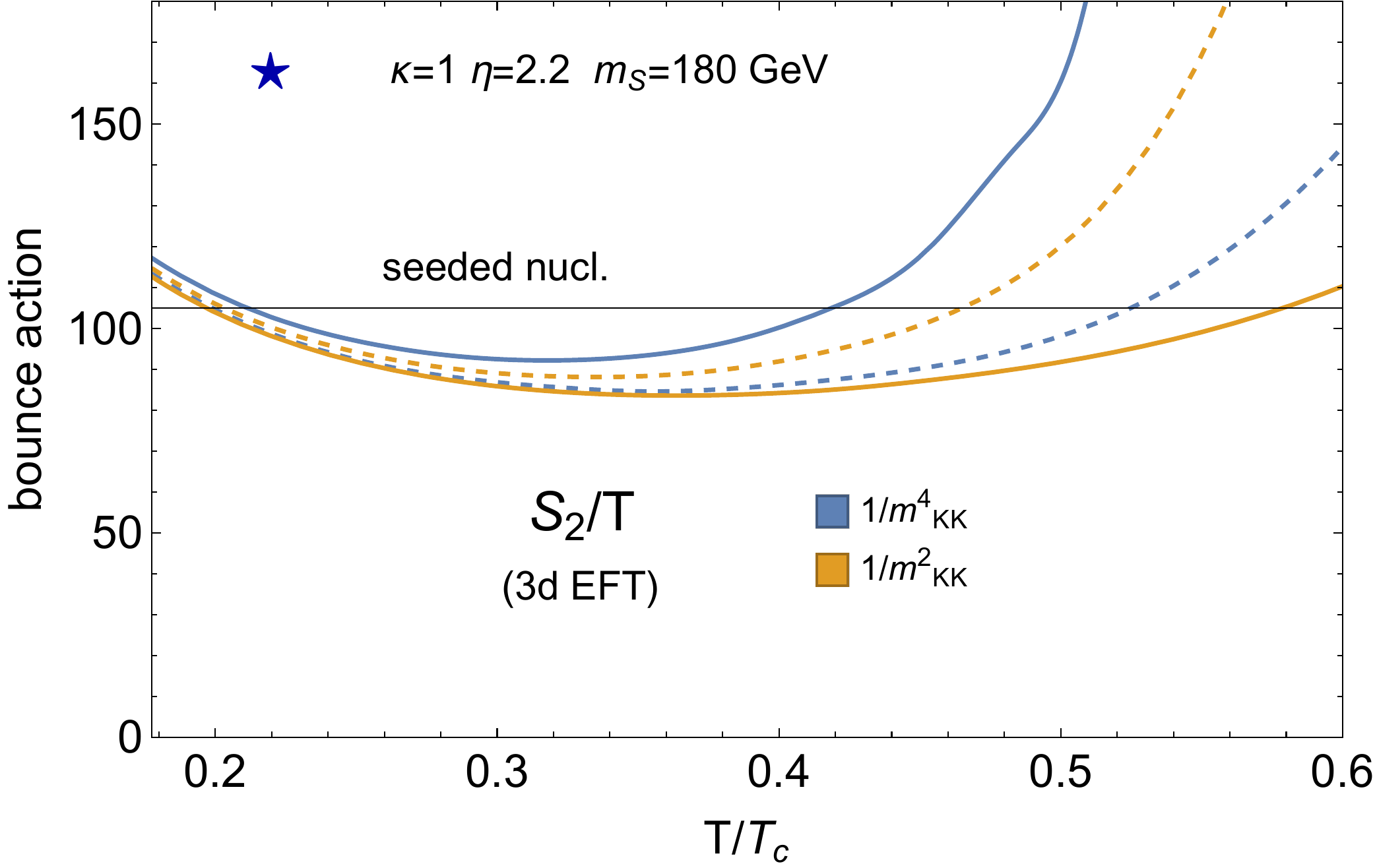}
 \caption{Benchmark point for
 $m_s = 180$ GeV with 
 $\kappa=1$ and $\eta=2.2$. 
 The horizontal gray line indicates the nucleation condition
 for a phase transition catalyzed by the domain walls.
 The solid lines show the bounce action $S_2/T$ as obtained within
 the effective 3d theory at $\mathcal{O}(1/m_\text{KK}^4)$ (blue)
 and $\mathcal{O}(1/m_\text{KK}^2)$ (orange). Dashed lines are shown
 as a further estimate of the theoretical uncertainty as explained
 in the text.
 The approximation scheme in which the continuum states are neglected
 is unable to identify the metastability of the false vacuum
 and no prediction for the tunneling can be obtained.}
 \label{fig:ben180}
\end{figure}

Our results for $S_2/T$ are presented
in Fig.\,\ref{fig:ben180}.
In the simplest approximation that neglects the continuum
KK states, the metastability of the 3d vacuum $(0,0)_\text{3d}$
is not captured at all, and thus no prediction for the tunneling
can be obtained (unlike for the benchmark point
shown in Fig.\,\ref{fig:ben250}, where the vanishing
of the barrier at the rolling temperature ensures a range
of validity for this approach).

The solid lines in Fig.\,\ref{fig:ben180} are obtained according
to the effective 3d potential at $\mathcal{O}(1/m^4_\text{KK})$
(blue) and $\mathcal{O}(1/m^2_\text{KK})$ (orange).
The dashed lines are similarly based on integrating
out the KK states with the corresponding accuracy. The
actual effective potential for $h_0$ and $s_0$ is however 
re--evaluated by substituting the
profiles of the KK states obtained as a function of $h_0, s_0$
back in the 4d potential. The numerical integration over $z$ then gives
an effective 3d potential that contains higher--order terms.
The discrepancy between the dashed and solid lines can then 
be used as a further estimate of the theoretical uncertainty.

As we can see from Fig.\,\ref{fig:ben180}, all the different
levels of approximation agree at lower temperatures
as the release point is there closer to the origin.
This is already enough to conclude that the phase transition
will certainly be completed as there is a range of temperatures
for which $S_2/T$ is below the nucleation
condition with small uncertainty. Points
belonging to the blue regions in the scan plots of Fig.\,\ref{fig:scan180}
and Fig.\,\ref{fig:scan250} are in fact conservatively 
determined in this way.

At higher temperatures, the release point moves further away from the 
origin and the theoretical uncertainty increases. We can nonetheless 
estimate the nucleation to be at $T \sim 0.5 \,T_c$ according to 
the dashed blue line, 
and look at the bounce
solution at this temperature to illustrate the general
shape of the seeded bubbles.

\begin{figure}
\centering
 \includegraphics[scale=0.463]{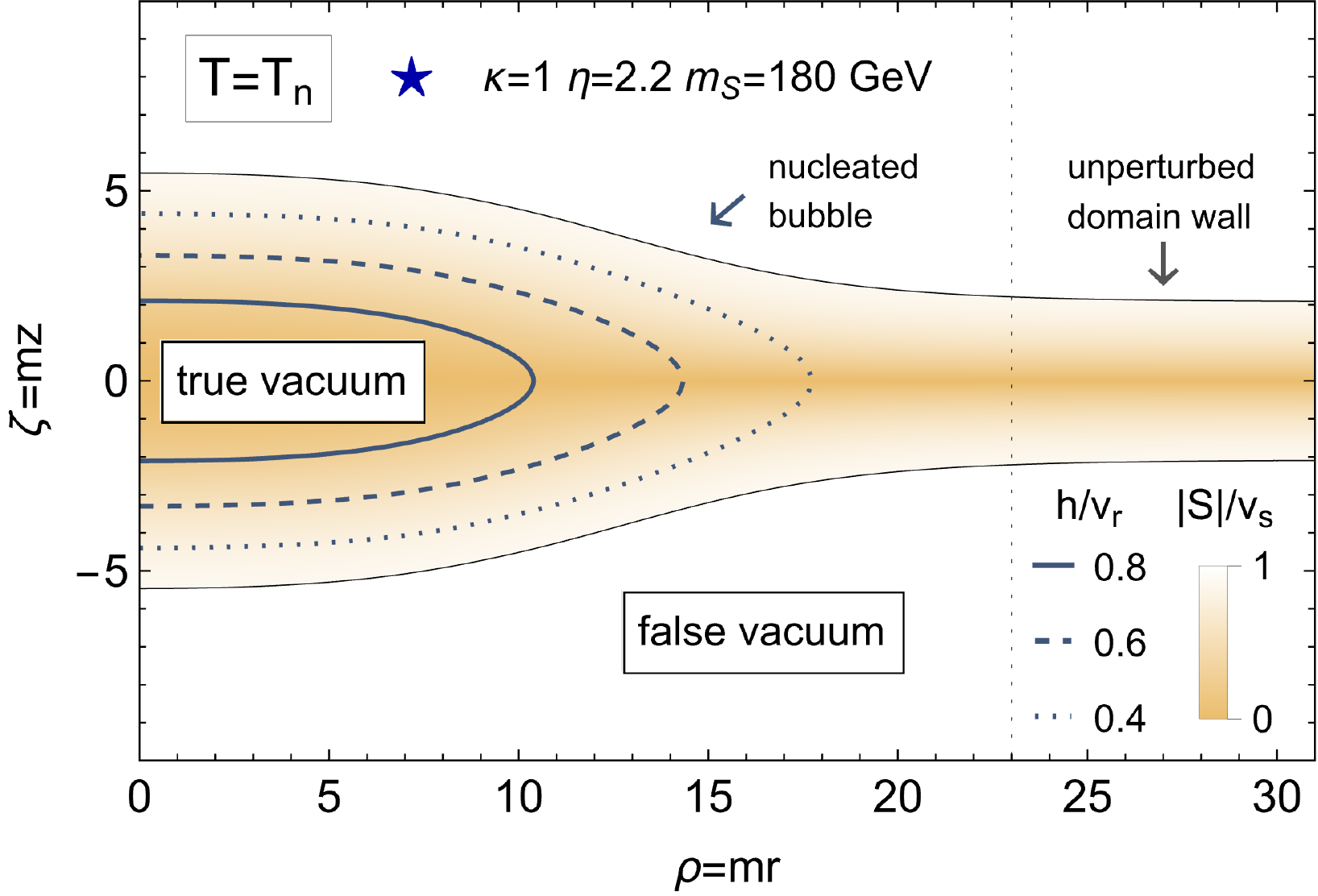}
 \caption{The shape of a bubble of true vacuum at the nucleation
 temperature in the benchmark
 point with $m_S = 180$ GeV, $\kappa=1$ and $\eta =2.2$
 as a function of the natural variables $\rho = m r$ and $\zeta = m z$.
 The Higgs field is shown as contour levels
 where it takes values $0.8$ (solid), $0.6$ (dashed) and 
 $0.4$ (dotted) times the vev at the center of the bubble,
 $v_r \simeq 200$ GeV at $(r=0,z=0)$. The singlet field normalized
 to $v_s \simeq 110$ GeV is shown according to the color 
 code in orange. The unperturbed domain wall profile at 
 $\rho \rightarrow \infty$ is modified to host a non vanishing
 Higgs vev at $\rho=0$. For $\zeta \rightarrow \pm 
 \infty$ the false vacuum $(h=0, S=\pm v_s)$ is approached everywhere.
 }
 \label{fig:bubble}
\end{figure}

The Higgs and singlet bounce profiles  
at the estimated nucleation temperature are
shown in Fig.\,\ref{fig:bubble} as contour plots
in the $(\rho,\zeta)$ plane, where $\rho = m r$ is the polar coordinate
in the domain wall plane rescaled by the inverse of the domain wall width,
and similarly for $\zeta = m z$ along the orthogonal direction.
The Higgs profile is shown by three contour levels 
where it is $0.8$ (solid), $0.6$ (dashed)
and $0.4$ (dotted) times the value that it takes at the center of the bubble
$(r=0,z=0)$, denoted by $v_r \simeq 200$ GeV.
The singlet profile is normalized to its vev in the false vacuum, $v_s
\simeq 110$ GeV, and is shown as a density plot according to the 
orange color code. 

Far from the bubble for $\rho\rightarrow \infty$,
the unperturbed, straight, domain wall interpolating
between $S=\pm v_s$ is recovered, and the Higgs field takes
a negligible vev. Similarly for $z \rightarrow \pm \infty$
the system
approaches the false vacuum $(0,\pm v_s)$, in agreement
with the boundary conditions in Sec.\,\ref{sec:svsuns}.

Around $\rho,\zeta=0$, we can see that a bubble of true vacuum
has been nucleated with the original
domain wall profile rearranging to host
a non--vanishing Higgs vev. 
We also note that the shape of the bubble
resembles more the one of a disk, or of a spheroid,
rather than a sphere, with the 
size in the domain wall plane being $\sim 4$ times
larger than the one along the orthogonal direction.

\section{Discussion}
\label{sec:discussion}

Our results in Sec.\,\ref{sec:results} 
show that domain walls can act
as local impurities at the time of
electroweak symmetry breaking, and can thus
have a strong impact on the dynamics
of the phase transition.
Interestingly, all the ingredients
for this to occur are already present in the minimal extension
of the SM including a $Z_2$--odd real scalar singlet
undergoing a two--step phase transition.
The tunneling process catalyzed by the domain walls
is studied 
with a new method based on dimensional
reduction, by which the seeded phase transition 
is reduced to a standard homogeneous problem in 3d,
and in a complementary way by
employing the thin wall approximation.
Our conclusion is that in the parameter space
under consideration
the vacuum decay seeded by domain walls is always faster
than the homogeneous tunneling, and it
ultimately determines the phenomenological
properties of the 
phase transition. 
In addition, new regions
of parameter space where the fields would be trapped
in the false vacuum become now viable
as a result of the inhomogeneous tunneling. \\

We shall now discuss the phenomenological
implications of a catalyzed electroweak
phase transition.

First of all,
the gravitational wave signal in this scenario
will be mainly controlled by the parameter
$\xi$ counting the number of defects
per Hubble volume\,\cite{PhysRevD.30.272}.
This in fact determines the duration of the phase transition
by setting the initial average distance between the
domain walls. In the case of rolling,
the walls will start dissociating as a whole
with velocity $v$
around $T_c$, and will then 
collide with an inverse time scale $\sim v \, \xi(T_c) H$.
If instead the transition proceeds by inhomogeneous
tunneling, bubbles will be nucleated on the domain walls
with a standard rate $\sim 100 H$
setting the time scale for the collisions
inside the domain walls.
 The collision with bubbles from the nearby walls
will on the other hand depend on the average
distance, thus giving an inverse time scale
$\sim v\,\xi(T_n) H$.
Characteristic features may then appear in the gravitational
wave spectrum associated to these two types of
collisions.

The amplitude
of the gravitational wave signal
from a catalyzed phase transition can then be significantly
larger than the one from its homogeneous counterpart
provided that $\xi< 100$, as this effectively 
corresponds to take a naturally smaller value 
for the inverse duration of the phase transition.
Two--step phase transitions in which there
is a larger gap in temperature between the steps
are more promising in this regard,
as domain walls will have more time to reach the scaling
regime where $\xi \sim \mathcal{O}(1)$. 
In addition, the sizeable violation of spherical
symmetry of the seeded bubbles can 
contribute to increase the production of gravitational
waves.
A quantitative analysis of these different
effects is left for future work. 

Let us now comment on how
electroweak baryogenesis may be
realized in this scenario in connection
with a possible origin of the $Z_2$ symmetry.
The singlet and the Higgs scalars 
may in fact arise as pseudo 
Nambu--Goldstone
bosons in the next--to--minimal composite 
Higgs model\,\cite{Gripaios:2009pe} based on
the $\text{SU}(4)/\,\text{Sp}(4)$ coset. 
The $Z_2$ symmetry comes then from the automorphisms of this coset
and can be identified with $\mathcal{CP}$ 
in the scalar sector,
with the Higgs and the singlet being even and odd, respectively.
Since a vacuum expectation value for $S$ breaks 
$\mathcal{CP}$ spontaneously, the model can lead
to successful electroweak baryogenesis
during the second step of the phase transition.
However, 
if the two possible
configurations $\langle S \rangle = \pm v_s$ are equally probable,
any asymmetry produced via this mechanism will eventually
average out when combining the contributions
of the two different domains. This issue was
circumvented in Refs.\,\cite{McDonald:1995hp,Espinosa:2011eu}
by adding a small explicit breaking of $\mathcal{CP}$
so that domain walls will collapse before the electroweak
phase transition, leaving the Universe with a single
homogeneous value of $v_s$ when baryogenesis
takes place.
Alternatively, the $Z_2$ symmetry
may never be restored at high temperatures elminating
altogether the formation of the
different domains\,\cite{Angelescu:2021pcd}.
However we remark that the presence of a small bias
induced by Planck--suppressed operators (as discussed
in Sec.\,\ref{sec:setup}) could prevent
the cancellation of the baryon asymmetry by favoring a slightly
larger volume corresponding to one minimum over the other,
as also noted in Ref.\,\cite{Espinosa:2011eu}, without necessarily
collapsing the domain walls at the time of the electroweak
phase transition.

Let us conclude this section by noticing that
defects, and in particular domain walls, can form and act as 
impurities also in models with a richer symmetry breaking pattern
than the one studied in this paper.
This may occur for instance in extensions of the SM with new electroweak
scalars\,\cite{Preskill:1991kd,Abel:1995wk,Chatterjee:2018znk,Eto:2018tnk},
but it can also be relevant for symmetry breaking
chains happening at higher energies in scenarios where
the SM gauge group is embedded in a larger group 
\cite{Yajnik:1998sw,Jeannerot:2003qv,Lazarides:2018aev,Chakrabortty:2019fov}.
We expect that the strategy developed here may be similarly
applied to these scenarios as well, possibly
involving other types of defects such as for instance 
in the case of cosmic strings.

\section*{Acknowledgements}
We are grateful to Riccardo Argurio and Diego Redigolo for 
useful discussions and comments on the draft.
We also thank Luigi Tizzano and Iason Baldes for useful discussions.

SB and AM are supported by the Strategic Research Program High-Energy Physics and the Research Council of the Vrije Universiteit Brussel, and by the ``Excellence of Science - EOS" - be.h project n.30820817.

\appendix
\section{Bubble tension in the thin wall approximation}
\label{app:bubble_tension}
In this Appendix we discuss how we can estimate the tension 
of the bubble in the thin wall limit for a multi-field potential.
The tension in the thin wall approximation is the one--dimensional bounce action evaluated across the bubble wall
\cite{Coleman:1977py,Linde:1981zj,Anderson:1991zb}
\be
S_1 = 
 \int_{R - \delta R}^{R+\delta R} \sum_i \frac{1}{2} 
 \left(\frac{\text{d} \phi_i}{\text{d} r} \right)^2 +V(\phi_i) \text{d}r,
\ee
where we will normalize $V(\phi_i^{f})=0$ for simplicity.
Using the equation of motion in the thin wall approximation, one can simplify this expression and get
\be
S_1 = 
 \int_{R - \delta R}^{R+\delta R}  2 V(\phi_i) ~\text{d}r.
\ee
Since the expressions for the bounce profiles $\phi_i$ are not known,
it is useful to make a change of variables in the integral,
moving from the radius to the fields themselves as done in the standard 
one-field case.
In multi-field potentials, this is complicated by the fact that 
we do not know the path in field space that minimizes the tension.
We shall here consider as optimal path the one that minimizes the scalar potential $V(\phi_i)$. Specializing to the two field case (having in mind the application
to the Higgs--singlet model of the main text), we can parametrize this path by using $\phi_1$ and defining a function
\be
\phi_2(\phi_1) \qquad \text{with} \qquad \phi_2(\phi_1^f) = \phi_2^f \qquad \phi_2(\phi_1^t) = \phi_2^t
\ee
that follows the path that minimizes the scalar potential. The function $\phi_2(\phi_1)$
 can be easily found numerically.
We can then make the change of variable from $r$ to $\phi_1$ and find the Jacobian of the transformation 
by using again the equation of motions in the thin wall approximation. We then obtain
\be
S_1 = 
 \int_{\phi_1^*}^{\phi_1^{f}} \sqrt{1+\left(\frac{d \phi_2}{d\phi_1}\right)^2} \sqrt{2 V(\phi_1,\phi_2(\phi_1))} d \phi_1.
\ee
In order to minimize the tension, the lower extremum of the integral is defined as the closest point to $\phi_1^f$ along 
the path such that  $V(\phi_1^*,\phi_2(\phi_1^*))=V(\phi_1^f,\phi_2^f)$.
As expected, this expression reduces to the standard thin wall tension for one-field if $\phi_2$ is 
actually a constant along the path.

\section{Estimate of the nucleation temperature}
\label{app:estimate}

\begin{figure}
\centering
 \includegraphics[scale=0.48]{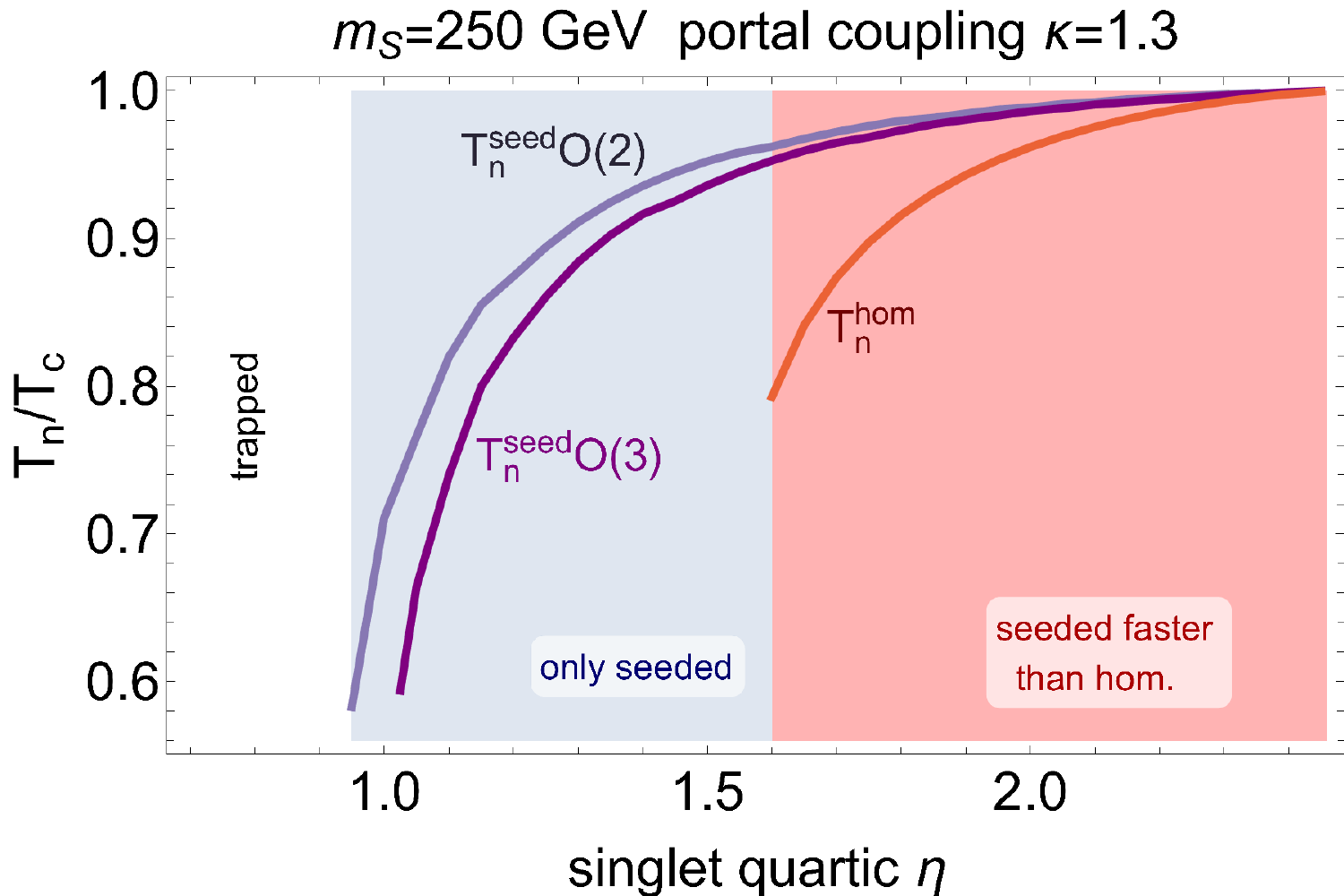}
 \caption{Nucleation temperature normalized
 to the critical temperature in the slice of parameter space obtained by
 fixing $m_S=250$ GeV and $\kappa = 1.3$. The red line shows the
 nucleation temperature of the homogeneous
 phase transition, $T_n^\text{hom}$, 
 computed numerically with \texttt{FindBounce}.
 The inhomogeneous nucleation temperature in the thin wall approximation
 is shown according to the $O(3)$ and $O(2)$ ansatz by the purple and light purple line,
 respectively. The color shading of the regions has the same meaning
 as in Fig.\,\ref{fig:scan250}.
 \label{fig:thinwall}
 } 
\end{figure}

In this Appendix
we push the thin wall approximation to its limit of validity and 
estimate the nucleation temperature moving more and more away from $T_c$.
We consider a vertical slice of the parameter space 
explored in Fig.\,\ref{fig:scan250} by fixing
$\kappa = 1.3$ while varying $\eta$.
For each value of $\eta$ we compute the $O(3)$ and $O(2)$ thin
wall bounce action as described in Section \ref{sec:TW},
and we derive the corresponding nucleation temperature
by the condition \eqref{eq:nucDW}.
Our results are shown in Fig.\,\ref{fig:thinwall},
where we have also included the homogeneous
nucleation temperature $T_n^\text{hom}$ 
evaluated with \texttt{FindBounce} for comparison.

As we can see,
the $O(2)$ and $O(3)$ approximations give similar results close to 
$T_c$, with the $O(2)$ ansatz providing in general a higher nucleation
temperature (or equivalently a smaller action).
The behaviour shown in Fig.\,\ref{fig:thinwall} 
agrees with the picture that the seeded phase transition
on the domain walls anticipates the wuould--be homogeneous
nucleation, as $T_n^\text{seed} > T_n^\text{hom}$
for all the relevant values of $\eta$.

\bibliographystyle{JHEP}
\bibliography{refs}

\end{document}